# The perovskite/transport layer interfaces dominate non-radiative recombination in efficient perovskite solar cells

Martin Stolterfoht[1,*], Pietro Caprioglio[1,3], Christian M. Wolff[1], José A. Márquez[2], Joleik Nordmann[1], Shanshan Zhang[1], Daniel Rothhardt[1], Ulrich Hörmann[1], Alex Redinger[2], Lukas Kegelmann[3], Steve Albrecht[3], Thomas Kirchartz[4], Michael Saliba[5], Thomas Unold[2,*], Dieter Neher[1,*]

[1]Institute of Physics and Astronomy, University of Potsdam, Karl-Liebknecht-Str. 24-25, D-14476 Potsdam-Golm, Germany.

[2]Department of Structure and Dynamics of Energy Materials, Helmholtz-Zentrum-Berlin, Hahn-Meitner-Platz 1, D-14109 Berlin, Germany

[3]Young Investigator Group Perovskite Tandem Solar Cells, Helmholtz-Zentrum Berlin für Materialien und Energie GmbH, Keleléstraße 5, 12489 Berlin, Germany

[4]Institut für Energie- und Klimaforschung, Forschungszentrum Jülich GmbH, 52425 Jülich, Germany

[5]Soft Matter Physics, Adolphe Merkle Institute, CH-1700 Fribourg, Switzerland

E-mail: stolterf@uni-potsdam.de, unold@helmholtz-berlin.de, neher@uni-potsdam.de

**Abstract**

Charge transport layers (CTLs) are key components of diffusion controlled perovskite solar cells, however, they can induce additional non-radiative recombination pathways which limit the open circuit voltage ($V_{OC}$) of the cell. In order to realize the full thermodynamic potential of the perovskite absorber, both the electron and hole transport layer (ETL/HTL) need to be as selective as possible. By measuring the quasi-Fermi level splitting (QFLS) of perovskite/CTL heterojunctions, we quantify the non-radiative interfacial recombination current for a wide range of commonly used CTLs, including various hole-transporting polymers, spiro-OMeTAD, metal oxides and fullerenes. We find that all studied CTLs limit the $V_{OC}$ by inducing an additional non-radiative recombination current that is significantly larger than the loss in the neat perovskite and that the least-selective interface sets the upper limit for the $V_{OC}$ of the device. The results also show that the $V_{OC}$ equals the internal QFLS in the absorber layer of (*pin*, *nip*) cells with selective CTLs and power conversion efficiencies of up to 21.4%. However, in case of less selective CTLs, the $V_{OC}$ is substantially lower than the QFLS which indicates additional losses at the contacts and/or interfaces. The findings are corroborated by rigorous device simulations which outline several important considerations to maximize the $V_{OC}$. This work shows that the real challenge to supress non-radiative recombination losses in perovskite cells on their way to the radiative limit lies in the suppression of carrier recombination at the perovskite/CTL interfaces.

**Introduction**

Huge endeavours are devoted worldwide to understanding and improving the performance of perovskite solar cells, which continue to develop at a rapid pace already outperforming other conventional thin-film technologies on small cells (< 1cm$^2$).[1] It is well established that further improvements will require suppression of non-radiative recombination losses to reach the full thermodynamic potential in terms of open circuit voltage ($V_{OC}$) and fill factor (FF).[2] As such, a major focus of the entire field to push the technology forward is targeted at reducing defect recombination in the perovskite bulk with numerous works highlighting the importance of grain boundaries in determining the efficiency losses.[3,4] In contrast, many other studies highlight the significance of traps at the perovskite surface which is likely chemically distinct from the bulk.[4–6] In many cases, performance improvements were achieved by mixing additives into the precursor solution including multiple cations and/or halides.[6–9] Interestingly, in most studies, a slower transient photoluminescence (TRPL) decay is shown as the figure of merit to prove the suppressed trap-assisted recombination in the bulk while implying its positive impact on the overall device efficiency.[3,6,10] Significantly fewer publications have focused on the importance of non-radiative recombination of charges across the perovskite/CTL interface.[11–13] Until recently it has been challenging to pinpoint the origin of these free energy losses in complete cells, although there have been some studies with valuable insight.[11–15] Methods that have been employed to measure interfacial recombination in perovskite stacks include transient photoluminescence (TRPL)[13,16,17] or reflection spectroscopy (TRS),[14] transient microwave conductivity (TRMC),[15] transient photovoltage (TPV)[18] or impedance spectroscopy.[11,19] Whilst TRPL, TRS and TRMC exhibit in principle the required time resolution to unveil the kinetics of the interface and bulk recombination, the interpretation of these transient measurements can be very challenging. The reasons are related to the inherent fact that extraction and recombination can both reduce the emitting species in the bulk, thus causing the signal decay.[2] While the correct interpretation of TRPL signals, for instance, is topic of ongoing discussions, several recent studies found the low-fluence signal decay to be primarily dominated by interfacial recombination[13,20] rather than extraction (diffusion) of charges to the transport layers.[21] Previously, a much more direct approach to decouple the origin of these recombination losses at each individual interface has been introduced based on absolute photoluminescence imaging.[20,22–24] In one of our recent studies, we have demonstrated how the device $V_{OC}$ can be explained through QFLS losses in the perovskite bulk and at the individual interfaces. Although the losses at the interfaces could be partially mitigated through the addition of interlayers between the CTLs and the absorber, they remained the limiting factor in our optimized cells. Whether this is, however, generic to different perovskite solar cell architectures and geometries remains an important question. In principal, there are two major groups of single junction perovskite solar cells depending on the arrangement of the electron/hole transport layer (ETL/HTL) and perovskite on the substrate: the "regular" *n*-illuminated *nip* configuration, and the "inverted" *p*-illuminated *pin* cell configuration. Despite the particularly simple device architecture of *pin* cells (e.g they require only a few nm of hole and electron transport layers without the need for chemical-doping and generally no high-temperature treatments are required during fabrication)[25–27] most research labs have adapted the *nip* configuration which is derived historically from dye sensitized solar cells and still deliver higher efficiencies overall.[7,28,29] In particular, the *nip* configuration showed traditionally higher open-circuit voltages (> 1.23 V),[7,11] although very recently > 1.2 V have been demonstrated also for *pin* type cells with a bandgap close to 1.6 eV.[30] Although the exact origin of this efficiency gap remains unclear today, different explanations have been put forward to explain this discrepancy, such as an improved crystallinity of the absorber when deposited on TiO$_2$ *nip* cells as well as superior properties of TiO$_2$ which has a near ideal large bandgap with a high carrier mobility (compared to C$_{60}$ for instance).[31,32]

In this work, we studied interfacial recombination losses by means of absolute PL measurements for a range of popular CTLs for "regular" (*nip*) and "inverted" (*pin*) perovskite solar cells including metal oxides, conjugated polymers, small molecules, and fullerenes. In particular, we also aim to compare the selectivity of ETLs and HTLs used for *nip* and *pin* configurations; i.e. for instance TiO$_2$ or SnO$_2$ vs. C$_{60}$, or doped Spiro-OMeTAD vs. PTAA. Here, we define the selectivity of a CTL as its ability to maintain the QFLS of the absorber while providing efficient majority carrier extraction. The results suggest that all studied CTL in the junction with the perovskite yield a substantially lower QFLS compared to the neat material on fused silica. Comparing the QFLS obtained on HTL/perovskite and perovskite/ETL heterojunctions with the *nip* or *pin* stacks suggests the validity of a simple superposition principle of non-radiative recombination currents at each individual interface. This also implies that the inferior interface dominates the energy loss in the final stack. We also find that cells with less selective contacts have a $V_{OC}$ that is substantially lower than the QFLS of the heterojunction. Extensive drift diffusion simulations identify reasons for this mismatched QFLS-$V_{OC}$ and highlight several important aspects for achieving high open-circuit voltages, including the role of energy level offsets and built-in voltage. By these means the presented results highlight that the primary non-radiative recombination loss channel of current perovskite cells is interfacial recombination at (or across) the perovskite/CTL interface, and that suppression of defect recombination in the perovskite bulk is an important yet secondary consideration today. This work represents a significant step forward towards understanding energy losses and the QFLS across the perovskite solar cells stack.

**Materials**

The studied CTLs in this work belong to 3 material classes, hole-transporting conjugated polymers, small molecules (either electron or hole transporting) and electron-transporting transparent metal oxides. The first class of materials, conjugated polymers, are currently attracting enormous attention due to their excellent film-forming abilities, moisture resistance, tunable energy levels and high charge selectivity.[6,33] Here, we studied highly selective wide-band gap donors such as PolyTPD and PTAA. In fact, the polymer PTAA is has been employed as capping hole-transporting layer in perovskite cells exceeding 22% efficiency.[6] In order to draw correlations between the QFLS and the energetics of the HTL, we also investigated other common organic semiconductors, namely P3HT (an intrinsic polythiophene-based macromolecular semiconductor which exhibits a considerably smaller bandgap and thus absorbs light throughout the visible spectrum),[34,35] as well as PEDOT:PSS (a nearly transparent and highly conductive composite comprising a highly *p*-doped derivative of polythiophene).[12] As small molecule HTL, we tested Spiro-OMeTAD which is arguably the most common HTL in *nip* perovskite solar cells.[36,37] Importantly, Spiro-OMeTAD requires external doping by different ionic salts and other additives,[37] which has been linked to device degradation and enhanced interface recombination.[11,36,38] For the case of small molecule ETLs, we tested the fullerene $C_{60}$ (with and without the interlayer LiF[20]) and the solution processable fullerene derivative PCBM.[27,39] Notably, both have been employed in high efficiency (> 20% PCE) *pin* cells.[27,39] Lastly, we studied the commonly used transparent metal oxides $TiO_2$ which is widely considered as an ideal electron transporting layer due to its high selectivity and high charge carrier mobility,[31] as well as $SnO_2$ - the preferred platform for planar efficient *nip* cells.[11] These materials are summarized in **Figure 1** along with their energy levels as measured by photoelectron spectroscopy in air.[40]

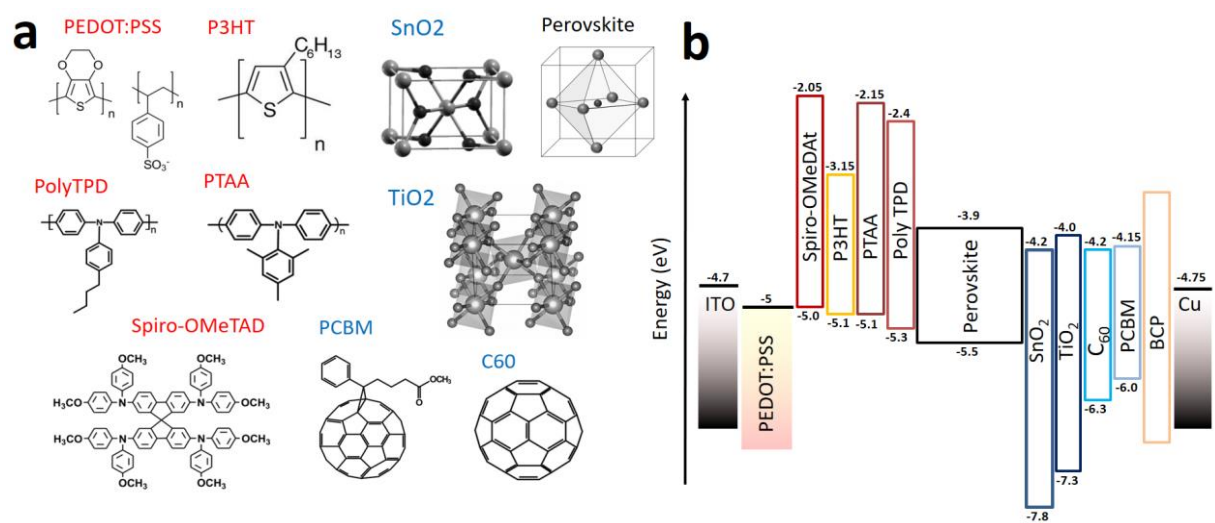

*Figure 1. Energy levels of the studied materials. The ionization potentials (IPs) were measured with photoelectron spectroscopy in air while the optical bandgaps were estimated from Tauc plots based on UVVis measurements. The IPs of $C_{60}$, $TiO_2$, and $SnO_2$ were outside the measurement range of the spectrometer (< -6.5 eV), therefore IPs previously determined from ultraviolet photon electron spectroscopy are plotted.[40] It is important to note that the plotted energy levels are only relevant for each film in isolation and by no means represent the true energetics in the complete solar cell stack where junctions form and the vacuum level may not constant across all interfaces.*

**Results**

**Comparison of CTLs for *pin* and *nip* type devices**

In order to quantify the free energy losses at the CTL/perovskite interface, we measured the absolute photoluminescence (PL) yield of heterojunctions containing the HTL (or ETL) adjacent to the perovskite, *pin* (*nip*) stacks, and also of complete cells including the metal electrodes. The absolute PL is a direct measure of the quasi-Fermi level splitting (QFLS or $\mu$) in the absorber,[41–45] and this approach has been recently applied to perovskite solar cells by various groups.[20,22–24] The ratio of emitted ($\phi_{em}$) and absorbed photon fluxes ($\phi_{abs}$) defines the absolute external PL quantum yield (PLQY):

$$\text{PLQY} = \frac{\phi_{em}}{\phi_{abs}} = \frac{J_{rad}/e}{J_G/e} = \frac{J_{rad}}{J_{R,tot}} = \frac{J_{rad}}{J_{rad} + J_{non-rad}} = \frac{J_{rad}}{J_{rad} + J_B + J_{p-i} + J_{i-n} + \ldots} \quad \text{(eq. 1)}$$

If all emission is from the direct recombination of free charges, and also every absorbed photon generates a free electron-hole pair, the PLQY also equals the ratio of the radiative recombination current density ($J_{rad}$)[41] and the total free charge generation current density ($J_G$). At $V_{OC}$, charge extraction is zero, meaning that the PLQY describes the ratio of $J_{rad}$ to the total recombination current ($J_{R,tot}$) of radiative and non-radiative losses ($J_{rad} + J_{non-rad}$), and that $J_{R,tot}$ is equal to $J_G$. Furthermore, $J_{non-rad}$ is equal to the sum of all non-radiative recombination pathways in the bulk ($J_B$), at the HTL/perovskite ($J_{p-i}$) and perovskite/ETL ($J_{i-n}$) interfaces, and potentially other losses (e.g. recombination in the transport layers, or at the CTL/metal interfaces). Using the expression for the radiative recombination current density according to Shockley-Queisser[41] and **Equation 1** we can write the QFLS as a function of the radiative efficiency

$$J_{rad} = J_{o,rad} e^{\mu/kT} \rightarrow \mu = kT \ln\left(\frac{J_{rad}}{J_{0,rad}}\right) = kT \ln\left(\text{PLQY}(\mu)\frac{J_G}{J_{0,rad}}\right) = kT \ln\left(\frac{J_G}{J_0}\right) \quad \text{(eq. 2)}$$

where $J_{o,rad}$ and $J_o$ are the radiative and the total thermal equilibrium recombination current densities in the dark. We note, that the PLQY depends itself on external conditions such as the illumination intensity or the internal QFLS. This originates from the fact that the non-radiative recombination pathways depend differently on the actual number of charge pairs present in the device compared to radiative recombination.[46] Thus, in order to predict the QFLS under 1 sun and open-circuit, the PLQY needs to be measured under the same illumination conditions. We also note that **Equation 2** shows that the QFLS is proportional to the logarithm of the PLQY which itself is limited by the largest recombination current in the denominator in **Equation 1**. In order to quantify the QFLS, the generated current density under illumination ($J_G$) and in the dark ($J_{0,rad}$) needs to be known, as well as the thermal energy (we measured a temperature of ~26-28°C on the sample under 1 sun equivalent illumination using a digital standard infrared sensor). The dark (and light) generation currents are obtained from the product of the $EQE_{PV}$ and the 300 K - black body (the solar) spectrum, respectively.[41,42,47,48] As such, we obtained a $J_{0,rad}$ of ~6.5x10$^{-21}$ A/cm$^2$ (±1x10$^{-21}$ A/cm$^2$) independent of the bottom CTL (**Supplementary Figure S1**) as it is predominantly determined by the tail absorption of the triple cation perovskite absorber layer (with Urbach energies around 15 meV). In all cases, the QFLS was measured by illuminating the films through the perovskite (or the transparent layer in case of *pin* or *nip* stacks) in order to avoid parasitic absorption of the studied CTL which can influence the QLFS if the parasitic absorption is significant (and $\phi_{abs}$ doesn't equal $J_G/e$ anymore). For instance, illuminating a perovskite/C$_{60}$ film through C$_{60}$ using a 445 nm laser causes a slightly lower QFLS compared to illuminating through the perovskite from the bottom (**Supplementary Figure S2**). The results of the PL measurements of the different transport layers are summarized in **Table 1** and plotted in **Figure 2**. All films, except samples containing Spiro-OMeTAD which required oxygen doping, were encapsulated around the corners of the substrate such that the laser (with spot size of ~1cm$^2$) did not directly hit the encapsulation glue. The films were measured directly after fabrication (see **Methods**) and all results were obtained as an average of multiple fabricated films (around 5 films for each individual layer). The spread of the measured QFLS of each measured film are shown in **Supplementary Figure S3** together with representative PL spectra **Supplementary Figure S4**.

**Table 1.** Optoelectronic quality of several tested CTL/perovskite layer junctions.

| Film | Abs | PLQY | $J_{0,nr}$ [Am$^{-2}$] | QFLS [eV] |
|---|---|---|---|---|
| ITO/Pero | 0.839 | 2.0×10$^{-5}$ | 3.5×10$^{-16}$ | 1.060 |
| PEDOT/Pero | 0.854 | 7.5×10$^{-5}$ | 9.9×10$^{-17}$ | 1.092 |
| P3HT/Pero | 0.848 | 7.7×10$^{-4}$ | 1.0×10$^{-17}$ | 1.152 |
| Pero/Spiro-OMeTAD | 0.944 | 1.4×10$^{-3}$ | 4.6×10$^{-18}$ | 1.172 |
| PTAA/PFN/Pero | 0.852 | 5.1×10$^{-3}$ | 1.3×10$^{-18}$ | 1.204 |
| PolyTPD/PFN/Pero | 0.851 | 7.3×10$^{-3}$ | 1.1×10$^{-18}$ | 1.208 |
| Pero | 0.850 | 1.4×10$^{-2}$ | 4.6×10$^{-19}$ | 1.231 |
| SnO2/Pero | 0.854 | 5.9×10$^{-3}$ | 1.5×10$^{-18}$ | 1.201 |
| TiO2/Pero | 0.854 | 2.1×10$^{-3}$ | 3.2×10$^{-18}$ | 1.181 |
| Pero/PCBM | 0.934 | 5.7×10$^{-4}$ | 1.3×10$^{-17}$ | 1.145 |
| Pero/C60 | 0.927 | 3.8×10$^{-4}$ | 1.8×10$^{-17}$ | 1.137 |
| Pero/LiF/C60 | 0.892 | 1.3×10$^{-3}$ | 4.9×10$^{-18}$ | 1.170 |

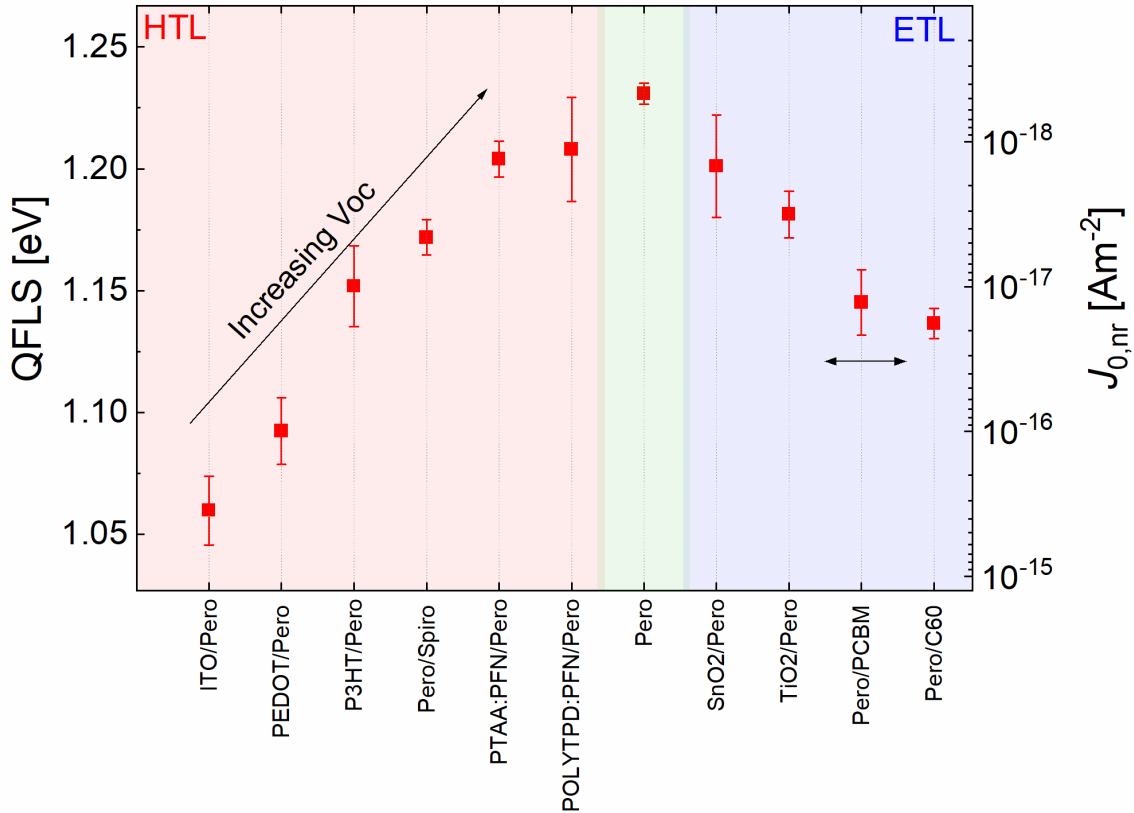

**Figure 2. The optoelectronic quality of charge transport layers.** *The calculated quasi-Fermi level splitting of the studied heterojunctions with different hole and electron transporting materials and of the neat absorber layer based on equation 2 using absolute photoluminescence measurements. The absorber was spin casted from the same solution for all transport layers.* The thermal non-radiative recombination current is plotted on the right and was obtained from $J_{0,\mathrm{nr}} = J_0 - J_{0,\mathrm{rad}}$.

**Figure 2** shows that the triple cation perovskite limits the QFLS to approximately 1.231 eV, which is ~110 meV below the radiative $V_{\mathrm{OC}}$ limit (where the PLQY equals 1). We note that we cannot rule out that this value is limited by recombination at the fused silica/perovskite interface, and thus the potential QFLS of the bare perovskite may be higher if it could be accessed without any underlying substrate. Consistent with this interpretation we observe a substantially lower QFLS (~40 meV) of the bare perovskite layer on a glass substrate (see **Supplementary Figure S5**). Recently significantly higher PLQY values above 20% were observed on methylammonium lead triiodide films where the top surface was passivated with tri-n-octylphosphine oxide (TOPO).[49] This results highlight the very high opto-electronic quality of the perovskite absorber comparable (or already better) than highly pure silicon or GaAs but also indicates substantial recombination losses at the perovskite surfaces. For the HTL/perovskite junctions we also tested the influence of the ITO, i.e. we compared the PLQY of perovskite layers on glass/HTL and on glass/ITO/HTL substrates, however this did not significantly influence the obtained QFLS within a small error except for samples with SnO$_2$ (see **Supplementary Figure S3**). Likewise, we tested the influence of the copper metal electrode on top of the C$_{60}$ in perovskite/C$_{60}$ heterojunctions and of *pin* stacks (**Supplementary Figure S6**). Overall, all these tests suggest that the relevant energy losses happen at the perovskite/CTL junctions and that there is an essentially lossless charge transfer between the metal electrodes and the HTL, which will be further discussed below. Interestingly, **Figure 2** shows that among all studied HTLs, the polymers PTAA/PFN and PolyTPD/PFN performed best - even outperforming the ubiquitous spiroOMeTAD - although we note that this may depend somewhat on the exact preparation conditions and may vary from lab to lab. Among the studied ETLs, SnO$_2$ and TiO$_2$ outperform the organic ETLs C$_{60}$ and PCBM which are usually used in *pin*-type cells. Therefore, this data suggests that the *p* interface is the limiting interface for *nip* cells, and the *n* interface for *pin* cells consistent with several earlier studies.[22] Moreover, we observe that the capping CTLs PCBM and $C_{60}$ are worse than Spiro-OMeTAD. Considering that the inferior interface will dominate the final $V_{OC}$ (**equation 3**), this might be one reason for the superior performance of *nip* cells today. An elegant approach to suppress non-radiative recombination at the perovskite/$C_{60}$ interface is to insert a thin LiF interlayer as demonstrated earlier (and in **Table 1**).[20]

A frequently arising question is how much the perovskite morphology, which potentially varies depending on the underlying CTL, could influence the obtained QFLS and the interpretation of the results. Thus, we performed top scanning electron microscopy and AFM measurements (see **Supplementary Figure S7**). Interestingly, we find the largest grains on a PEDOT bottom CTL despite it being the worst among the studied transport layers. The largest grain size distribution is visible on perovskite films on $TiO_2$ while the perovskite morphology on all other substrates appears, at least qualitatively, similar where we observe relatively small grains (< 10 – 100 nm). In addition, AFM measurements reveal root mean square surface roughnesses varying from 12 – 27 nm, where the perovskite on PolyTPD/PFN and PTAA/PFN appears to be roughest (> 20 nm) while the perovskite film on $TiO_2$ is the smoothest. Overall, considering these results it seems unlikely that the perovskite bulk morphology can explain the changes in the non-radiative recombination loss currents which increase by orders of magnitude depending on the underlying substrate (as shown in **Figure 2**). It is worth to note that these results do not allow distinguishing whether the critical recombination loss occurs across the perovskite/CTL interface, or at the perovskite surface next to the interface. In any case, the presence of the additional CTL triggers additional (non-radiative) interfacial recombination losses, which are dominating the non-radiative recombination losses.

**Comparison of the QFLS and device $V_{OC}$ and origin of free energy losses**

In the following, we aim to compare the non-radiative recombination losses at the *p*- and *n*-interfaces with the QLFS of the *pin* stacks and the $V_{OC}$ of the complete cells with $C_{60}$ as ETL. **Figure 3a** shows that the device $V_{OC}$ (black dots) generally increases with the average QFLS of the *pin*-stack (orange stars) which was taken as an average as obtained on 3-4 stacks for each configuration. Importantly, for optimized cells with highly-selective HTLs such as PolyTPD or PTAA, the $V_{OC}$ matches the QFLS of the stack within a small error. On the other hand, in case of the less selective PEDOT and P3HT bottom layers, the $V_{OC}$ was found to be substantially lower than the corresponding QFLS. The current density vs. voltage characteristics corresponding to cells with different HTLs are shown in **Figure 3b** which highlight the large differences in the measured open-circuit voltages. Device statistics of individually measured stacks are shown in **Supplementary Figure S9**. We note that our devices with LiF/$C_{60}$ as ETL reach efficiencies of up to 21.4% with a $V_{OC}$ of ~1.2 V (for a perovskite with a bandgap of ~1.6 eV), which is among the highest reported values for *pin*-type cells (**Supplementary Figure S10**).[30]

As another important finding, we observe that the $V_{OC}$ of devices with the relatively selective PTAA/PFN and PolyTPD/PFN bottom layers and a $C_{60}$ ETL equals the QFLS of the less selective perovskite/$C_{60}$ junction (blue line), and that this is also nearly identical to the QFLS of the *pin* stack. This indicates that for these particular cells, the losses determining the $V_{OC}$ occur almost entirely at the inferior interface to the perovskite while the electrodes are not causing additional $V_{OC}$ losses. Consistent with this explanation is the fact that, under conditions where the dark injection current equals the generation current, the external electroluminescence quantum yield ($EQE_{EL}$ ~3×10$^{-4}$ for both devices) approaches the PLQY of the stack within a factor of two (5.9x10$^{-4}$ for PTAA and 4.6x10$^{-4}$ for PolyTPD), see **Figure 3c**. As such improving the perovskite/ETL interface, e.g. by inserting LiF between the perovskite and $C_{60}$ allows to improve both the QFLS of the *pin* stack and the $V_{OC}$ to 1.17 V on average (with a PLQY of ~1.3 x10$^{-3}$ and $EQE_{EL}$ of ~8.3x10$^{-4}$).[20] Similarly, for devices with PEDOT, the inferior interface (PEDOT/perovskite) limits the QFLS of the stack, however, there are additional energy losses between the QFLS and the $V_{OC}$ indicating additional losses at the contacts and/or interfaces. This important point will be addressed below. Again, we note that the measured $EQE_{EL}$ (~1.4x10$^{-8}$) of the device roughly matches the expected value from the $JV$-scan (an $EQE_{EL}$ of 3.8x10$^{-8}$ is expected for a $V_{OC}$ of 0.9 V) but that this value is orders of magnitude lower than the PLQY of the stack (~1x10$^{-5}$). Lastly, films with P3HT lie somewhat in between PEDOT and PTAA (PolyTPD) devices. Here, both interfaces (P3HT/perovskite and perovskite/C60), appear to be equally limiting the QFLS of the stack which also lies below the QFLS of the individual heterojunctions. Moreover, like observed for PEDOT devices, the $V_{OC}$ is markedly smaller than the QFLS of the absorber suggesting additional losses, while we also observe a considerable mismatch between PLQY$_{STACK}$ (6.2x10$^{-5}$) and $EQE_{EL}$ (~9x10$^{-7}$). We note that the $EQE_{EL}$ is again very close to the $EQE_{EL}$ of 1.8x10$^{-6}$ that is expected for a P3HT device with a $V_{OC}$ of ~1.0 V.

Our optimized *nip*-cells (with the ETLs $SnO_2$ and $TiO_2$, and SpiroOMeTAD as the HTL) behave similarly to the optimized *pin*-cells with PTAA or PolyTPD, with a close match between the average device $V_{OC}$ (~1.15V) and the average internal QFLS (1.161 eV and 1.168 eV for $TiO_2$ and $SnO_2$ based cells, respectively) under 1 sun conditions. Considering the large QFLS potential of the neat perovskite of 1.23 eV, the data confirms our fundamental assertion that the perovskite/transport layer interfaces dominate the non-radiative recombination current in perovskite solar cells. All results obtained on *nip*-cells are shown in **Supplementary Figure S11**.

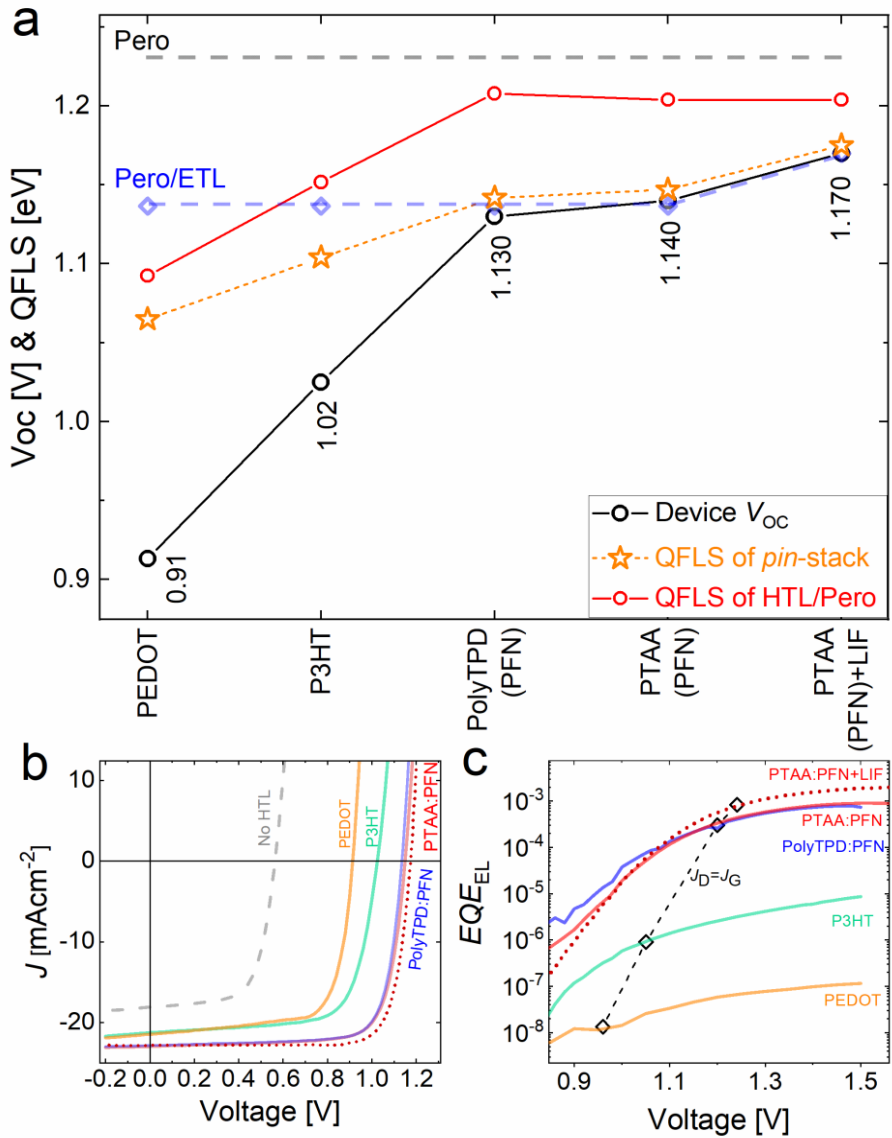

**Figure 3. Open-circuit voltage, quasi-Fermi level splitting and electroluminescence of *pin* cells.** *(a) Average $V_{OC}$ of pin cells employing different conjugated polymers as HTLs and a $C_{60}$ ETL, compared to the average QFLS of the corresponding HTL/perovskite bilayers (red), and of the pin stacks (orange). The QFLS of the perovskite/$C_{60}$ junction and of the neat perovskite on fused silica are shown in dashed blue and black lines, respectively. (b) Corresponding current density vs. voltage characteristics of the pin cells with different HTLs, and (c) the external electroluminescence efficiency as a function of voltage. The dashed line shows conditions where the dark injection and light generation currents are equal for each device.*

**Understanding the QFLS across the *pin* (*nip*) junction**

The experimental results in the previous sections show that the QFLS equals the device $V_{OC}$ in case of selective transport layers which also indicates that interfacial recombination in these devices lowers the QFLS throughout the whole bulk equally. However, if a less selective transport layer is employed such as PEDOT, or P3HT, then the device $V_{OC}$ is lower than the QFLS in the perovskite layer. In such cases, at least one QFL bends, presumably at the interfaces or contacts, causing a further reduction in the electrochemical potential of the photogenerated charges in a specific region of the multilayer device, and that this bending has a much larger effect on the final $V_{OC}$ than on the average QFLS in the perovskite bulk. In order to check whether this phenomenon depends on the charge carrier generation profile, we analysed all samples by illuminating the samples through the bottom glass or top using a 445 nm laser. However, we found the QFLS to be quite independent of the direction of illumination (< 20 meV difference) as long as the parasitic absorption of the CTL is not substantial (**Supplementary Figure S2**). We also illuminated full devices with light of different penetration depths, while ensuring the same generation current (or $J_{SC}$) through proper adjustment of the illumination intensity (**Supplementary Figure S12**). These measurements yielded the same $V_{OC}$. It can be concluded that neither the QFLS nor the $V_{OC}$ depends significantly on the charge generation profile, which we attribute to the rapid diffusion of charges through the perovskite. Considering perovskite mobilities on the order of ≈10 cm$^2$V$^{-1}$s$^{-1}$ (ref.[50]), within their bulk lifetime (~1000 ns)[20] carriers can diffuse through the absorber layer (with time constant $4d^2/\pi D^2$)[16,51] multiple times back and forth.

In order to understand the spatial distribution of the recombination losses and the QFLS, we simulated our perovskite solar cells using the well-established drift-diffusion simulator SCAPS.[52] These simulations take into account previously measured interface recombination velocities and perovskite bulk lifetimes.[20] As detailed in the previous work,[20] an interface recombination velocity ($S$) of around ~1000 cm/s was estimated at the perovskite/C$_{60}$ interface based on TRPL measurements, while the PTAA/PFN/perovskite interface recombination velocity was found to be smaller, around ~200 cm/s. The simulated electron/hole quasi-Fermi levels ($E_{F,e}$ and $E_{F,h}$) at $V_{OC}$ are shown along with the conduction and valence bands in **Figure 4a** for a PTAA/PFN/perovskite/C$_{60}$ device, while important simulation parameters listed in **Supplementary Table S1**. Qualitatively, these simulations confirm that $E_{F,e}$ and $E_{F,h}$ are spatially flat in the perovskite bulk and extend to the corresponding electrodes (cathode for electrons and anode for holes) which explains that $eV_{OC}$ is nearly identical to the QFLS (of ~1.13 eV) in these devices. Importantly, to reproduce the comparatively high open-circuit voltages (~1.14 V) and FFs (up to 80%) of these devices, a considerable built-in voltage ($V_{BI}$) of at least 1.0 V had to be assumed considering realistic interface recombination velocities. Otherwise, a strong backfield would hinder charge extraction in forward bias but also accumulate minority carriers at the wrong contact (**Supplementary Figure S13**). Moreover, we had to assume a small majority carrier band offset ($\Delta E_{maj} < 0.1$ eV) between the perovskite valance/conduction band and the HOMO/LUMO of the HTL/ETL, respectively. The reason is that such offsets would considerably increase the interfacial recombination loss through a large increase in the carrier density in the CTL (see **Supplementary Figure S14**). We note that such a near perfect band alignment is actually not really consistent with the energy levels as obtained from PESA on the individual layers (**Figure 1b**), however, it is important to note that the energetics could be very different in the composite solar cells stack due to the formation of junctions.

To simulate the *pin* stack with a PEDOT:PSS bottom layer (**Figure 4b**), we simplified the HTL by a metal with a work function of 5 eV, a high surface recombination velocity for holes and an intermediate value for electrons (**Supplementary Table S1**). We find that the hole-QFL bends indeed at the interface, giving rise to the observed QFLS-$V_{OC}$ mismatch. We acknowledge that these simulations only illustrate one possible scenario of the internal device energetics using a set of plausible parameters, and thus different energetic alignments cannot be excluded. In order to generalize the conditions under which the $V_{OC}$ deviates from the QFLS, we extended our simulations by studying a wide range of parameters (**Supplementary Table S1**). We found that at least two requirements must be fulfilled: a) a band offset for the majority carrier of at least ~0.2 eV, and b) a sufficiently high recombination velocity (> 1 cm/s), otherwise $E_{F,e}$ and $E_{F,h}$ can remain flat despite the energy offset (**Supplementary Figure S14**). Lastly, the device simulations also predict that the band offset for the minority carrier at the perovskite/CTL interfaces ($\Delta E_{min}$) is in principle not a decisive parameter in determining the recombination losses as long as $\Delta E_{min}$ is larger than only 0.1 eV which is further discussed at **Supplementary Figure S15**. Overall, we conclude that the defect density at the interfaces is the most critical parameter in determining the interface recombination velocity ($S$) and the non-radiative recombination losses, while a perfect energy level alignment of all layers is also a crucial requirement to maximize the $V_{OC}$.

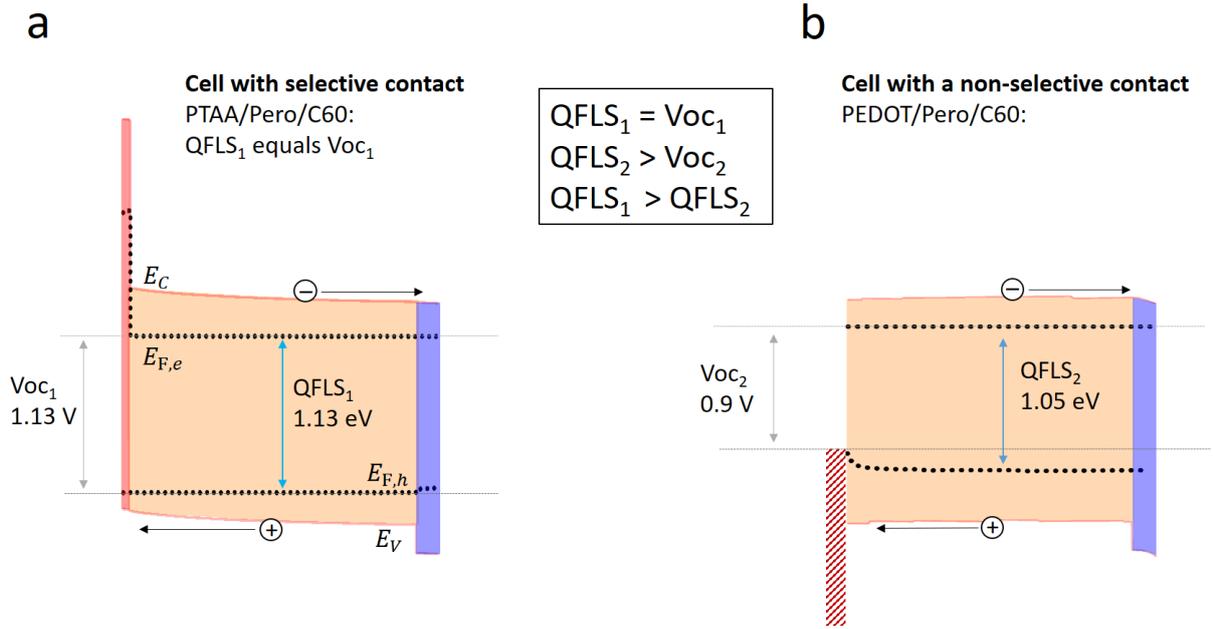

**Figure 4. Simulation of the QFLS and $V_{OC}$ of *pin*-type devices using SCAPS.** *The simulated quasi-Fermi level splitting (QFLS) in junctions with (**a**) selective transport layers (PTAA/perovskite/$C_{60}$) is identical to $eV_{OC}$ but not in case of non-selective (**b**) transport layers (PEDOT/perovskite/$C_{60}$) where the hole QFL bends at the interface to PEDOT. The perovskite is represented in orange showing unoccupied states in between the conduction band minimum ($E_C$) and valance band maximum ($E_V$), while the dashed lines show the electron and hole quasi-Fermi levels ($E_{F,e}$ and $E_{F,h}$), the resulting QFLS in the absorber and the open-circuit voltage ($V_{OC}$) at the contacts. PTAA (red) and $C_{60}$ (blue) are represented by their unoccupied states in between the highest and lowest unoccupied molecular orbitals. Occupied states are drawn for PEDOT in striped red.*

**Conclusions**

Using absolute PL measurements, we were able to decouple the origin of non-radiative recombination losses for cells in *pin* and *nip* configurations fabricated from different CTLs. We found that a range of the most common CTLs induce large non-radiative recombination currents which dwarf the non-radiative losses in the perovskite bulk. We identified that the most selective layers are the polymers PTAA and PolyTPD and $SnO_2$ which are outperforming the omnipresent $TiO_2$ and SpiroOMeTAD although we acknowledge that this can vary depending on the exact preparation conditions. Nevertheless, for *pin*-cells the perovskite/$C_{60}$ interface was found to be a major issue which induces more interfacial recombination than Spiro-OMeTAD or $TiO_2$ which could be one reason for the lower performance of *pin*-type cells with the standard electron transporter $C_{60}$. By comparing the QFLS of bilayers (with one CTL adjacent to the perovskite) and complete stacks with the device $V_{OC}$ shows that the relevant energy losses happen at the top interface in cells with relatively selective CTLs such as PTAA and PolyTPD, $SnO_2$ and $TiO_2$. In these cells, the electron/hole QFLs are expected to be spatially flat throughout the junction to the electrodes, meaning that the QFLS in the perovskite bulk determines the $V_{OC}$ of the cells. However, in cells with less selective HTLs such as PEDOT or P3HT, the $V_{OC}$ is lower than the QFLS in the absorber which indicates substantial losses at the contacts and/or interfaces. The fundamental study was validated in high-efficiency perovskite cells in *pin*-configuration with PCEs up to 21.4%. Lastly, device simulations were employed which substantiate the understanding obtained from these experimental results but also highlight the importance of a high built-in voltage and negligible energetic offsets between the perovskite and the transport layers. This contribution shows that the open-circuit voltage of perovskite cells can be well understood considering the interfacial recombination but also that the interfaces represent the biggest challenge in improving this technology to its radiative limit.

**Methods**

**Absolute Photoluminescence Imaging:** Excitation for the PL imaging measurements was performed with a 445 nm CW laser (Insaneware) through an optical fibre into an integrating sphere. The intensity of the laser was adjusted to a 1 sun equivalent intensity by illuminating a 1 cm² -size perovskite solar cell under short-circuit and matching the current density to the $J_{SC}$ under the sun simulator (22.0 mA/cm² at 100 mWcm$^{-2}$, or 1.375x10$^{21}$ photons m$^{-2}$s$^{-1}$). A second optical fiber was used from the output of the integrating sphere to an Andor SR393i-B spectrometer equipped with a silicon CCD camera (DU420A-BR-DD, iDus). The system was calibrated by using a calibrated halogen lamp with specified spectral irradiance, which was shone into to integrating sphere. A spectral correction factor was established to match the spectral output of the detector to the calibrated spectral irradiance of the lamp. The spectral photon density was obtained from the corrected detector signal (spectral irradiance) by division through the photon energy ($hf$), and the photon numbers of the excitation and emission obtained from numerical integration using Matlab. In a last step, three fluorescent test samples with high specified PLQY (~70%) supplied from Hamamatsu Photonics where measured where the specified value could be accurately reproduced within a small relative error of less than 5%. ***Measurement conditions:*** All films and cells were prepared fresh and immediately encapsulated in a glovebox after preparation with the exception of films and cells with spiroOMeTAD which require oxygen doping for enabling sufficient transport capability in the device (non-oxygygen treated spiroOMeTAD cells exhibited FFs below 20 % with negligible photovoltaic performance). Thus, films and cells with spiroOMeTAD were treated in atmosphere overnight at 25% relative humidity, and subsequently encapsulated before the PL measurements. The PL of the samples was readily recorded after mounting the sample after an exposure between 10-20 s to the laser light. Thus, the PLQY is obtained on timescales relevant to the $V_{OC}$ measurements on the cells. We note that all absolute PL measurements were performed on films with the same HTL, ETL and perovskite thicknesses as used in the operational solar cells. The absorption of the samples was considered in the PLQY calculation and was approximately 85% for cells illuminated through the top encapsulation glass, and ~93% through the bottom glass.

**Electroluminescence:** Absolute EL was measured with a calibrated Si photodetector (Newport) connected to a Keithley 485 pico Ampere meter. The detector (with an active area of ~2 cm²) was placed directly in front of the device (< 0.5 cm) and the total photon flux was evaluated considering the emission spectrum of the solar cell and the external quantum efficiency of the detector (around 86 % in the relevant spectral regime). A slight underestimation of the $EQE_{EL}$ (≈1.25x) cannot be excluded at present as some photons from the solar cells may escaped to the side and were not detected. A forward bias was applied to the cell using a Keithley 2400 source-meter and the injected current was monitored. Measurements were conducted with a home written LabVIEW routine. Typically, the voltage was increased in steps of 20 mV and the current stabilized for typical 1s at each step. No relevant changes in the $EQE_{EL}$ were observed for different stabilization times.

**Device Fabrication:** Pre-patterned 2.5x2.5cm² 15 Ω/sq. ITO (Automatic Research, Germany), glass or fused silica substrates were cleaned with acetone, 3% Hellmanex solution, DI-water and *iso*-propanol, by sonication for 10min in each solution. After a microwave plasma treatment (4 min., 200W), the samples were transferred to an N₂-filled glovebox (except PEDOT:PSS which was spincoated in air) where different CTLs were spincoated from solution.

***Bottom selective contacts: (HTLs or ETLs):*** PEDOT:PSS (Heraeus Celivious 4083) was spincoated at 2000 r.p.m for 40s (acceleration 2000 r.p.m/s) and subsequently annealed at 150 °C for 15 minutes; P3HT (Sigma Aldrich, Mn~27 000) was spincoated from a 3 mg/mL DCB solution at 3000 r.p.m for 30s (acceleration 3000 r.p.m/s) and subsequently annealed 100 °C for 10 minutes. P3HT films were also oxygen plasma treated for 5 s to ensure sufficient wetting of the perovskite as discussed in a previous work.[35] PolyTPD (Ossila) was spincoated from a 1.5 mg/mL DCB solution at 6000 r.p.m for 30 s (acceleration 2000 r.p.m/s) and subsequently annealed 100 °C for 10 minutes. PTAA (Sigma Aldrich) was spincoated was spincoated from a 1.5 mg/mL Toluene solution at 6000 r.p.m for 30 s (acceleration 2000 r.p.m/s) and subsequently annealed 100 °C for 10 minutes. For PTAA and PolyTPD coated samples, a 60 µL solution of PFN-P2 (0.5 mg/mL in methanol) was added onto the spinning substrate at 5000 rpm for 20 s resulting in a film with a thickness below the detection limit of our AFM (< 5 nm). For compact/mesoporous TiO2 samples, first a nippon Sheet Glass 10 Ω/sq was cleaned by sonication in 2% Hellmanex water solution for 30 minutes. After rinsing with deionised water and ethanol, the substrates were further cleaned with UV ozone treatment for 15 min. Then, 30 nm TiO₂ compact layer was deposited on FTO via spray pyrolysis at 450°C from a precursor solution of titanium diisopropoxide bis(acetylacetonate) in anhydrous ethanol. After the spraying, the substrates were left at 450°C for 45 min and left to cool down to room temperature. Then, a mesoporous TiO$_2$ layer was deposited by spin coating for 20 s at 4000 rpm with a ramp of 2000 rpm s-1, using 30 nm particle paste (Dyesol 30 NR-D) diluted in ethanol to achieve 150-200 nm thick layer. After the spin coating, the substrates were immediately dried at 100°C for 10 min and then sintered again at 450°C for 30 min under dry air flow. Before processing the perovskite layer TiO2 coated films were microwave plasma treatment (4 min., 200W). Compact SnO$_2$ films were fabricated by using a Tin(IV) oxide nanoparticle dispersion diluted 1:7 vol. with DI-H2O and filtered through 0.45 μm PVDF filter prior to spin coating on the substrate at 2000 rpm (acceleration 2000 r.p.m/s) for 30 s. After 20 minutes of annealing at 150 °C, the spin coating procedure was repeated and the samples were annealed again for 30 more minutes. Before processing the perovskite layer TiO2 coated films were microwave plasma treatment (4 min., 200W).

*Perovskite Layer:* The triple cation perovskite solution was prepared by mixing two 1.3 M FAPbI$_3$ and MAPbBr$_3$ perovskite solutions in DMF:DMSO (4:1) in a ratio of 83:17 which we call "MAFA" solution. The 1.3 M FAPbI$_3$ solution was thereby prepared by dissolving FAI (722 mg) and PbI$_2$ (2130 mg) in 2.8 mL DMF and 0.7 mL DMSO (note there is a 10% excess of PbI$_2$). The 1.3 M MAPbBr$_3$ solution was made by dissolving MABr (470 mg) and PbBr$_2$ (1696 mg) in 2.8 mL DMF and 0.7 mL DMSO (note there is a 10% excess of PbBr$_2$). Lastly, 40 $\mu L$ of a 1.2M CsI solution in DMSO (389 mg CsI in 1 mL DMSO) was mixed with 960 $\mu L$ of the MAFA solution resulting in a final perovskite stoichiometry of (CsPbI$_3$)$_{0.05}$[(FAPbI$_3$)$_{0.83}$(MAPbBr$_3$)$_{0.17}$]$_{0.95}$ in solution. The perovskite film was deposited by spin-coating at 4000 r.p.m (acceleration 1300 rpm/s) for 35 seconds; 10 Seconds after the start of the spinning process, the spinning substrate was washed with 300 μL EA for approximately 1 second (the anti-solvent was placed in the centre of the film). The perovskite film was then annealed at 100 °C for 1 hr on a preheated hotplate.

*Top selective contacts: (HTLs or ETLs):* SpiroOMeDAT was spincoated from a *spiro-OMeTAD* (Merck) solution in chlorobenzene (70 mM) at 4000 rpm for 20 s (acceleration 4000 rpm/s). Spiro-OMeTAD was doped with bis(trifluoromethylsulfonyl)imide lithium salt (Li-TFSI, Sigma-Aldrich), tris(2-(1H-pyrazol-1-yl)-4-tert-butylpyridine)-cobalt(III) tris(bis(trifluoromethylsulfonyl)imide) (FK209, Dynamo) and 4-tert-Butylpyridine (tBP, Sigma-Aldrich). The molar ratio of additives for spiro-OMeTAD was: 0.5, 0.03 and 3.3 for Li-TFSI, FK209 and tBP respectively. PC$_{61}$BM (Solenne BV) was spincoated from a 30 mg/mL DCB solution at 6000 rpm (acceleration 2000 r.p.m/s) for 30 s and the resulting Perovksite/PCBM film further annealed at 100 °C for 30 minutes. For C$_{60}$ (Creaphys) and LiF ETLs, the perovskite films were transferred to an evaporation chamber where 30 nm of C$_{60}$ (1 nm of LiF) were deposited at 0.1 Å/s (0.03 Å/s) under vacuum (p = 10$^{-7}$ mbar).

*PIN devices:* The cells were completed by transferring the samples to an evaporation chamber where 8 nm BCP (Sigma-Aldrich) at 0.2 A/s and 100 nm copper (Sigma-Aldrich) at 0.6 Å/s were deposited under vacuum (p = 10$^{-7}$ mbar).

*NIP devices:* The cells were completed by transferring the samples to an evaporation chamber where 100 nm gold (0.7 Å/s) were deposited under vacuum (p = 10$^{-7}$ mbar). *nip*-cells were oxgygen doped overnight at 20% relative humidity prior to device and PL measurements.

**Current density-voltage characteristics:** $JV$-curves were obtained in a 2-wire source-sense configuration with a Keithley 2400. An Oriel class AAA Xenon lamp-based sun simulator was used for illumination providing approximately 100 mW cm$^{-2}$ of AM1.5G irradiation and the intensity was monitored simultaneously with a Si photodiode. The exact illumination intensity was used for efficiency calculations, and the simulator was calibrated with a KG5 filtered silicon solar cell (certified by Fraunhofer ISE). The temperature of the cell was fixed to 25 °C and a voltage ramp of 67 mV/s was used. A spectral mismatch calculation was performed based on the spectral irradiance of the solar simulator, the EQE of the reference silicon solar cell and 3 typical EQEs of our cells. This resulted in 3 mismatch factors of $M$ = 0.9949, 0.9996 and 0.9976. Given the very small deviation from unity the measured $J_{SC}$ was not corrected by the factor $1/M$. All EQEs presented in this work were measured by ISE-Fraunhofer.

**SCAPS simulations:** Simulation parameters and further details are discussed at **Supplementary Table S1**.

**Acknowledgements.** We thank Lukas Fiedler and Frank Jaiser for lab assistance. Florian Dornack and Andreas Pucher for providing measurement and laboratory equipment. Philipp Tockhorn for characterization of $SnO_2$ based cells. This work was in part funded by HyPerCells (a joint graduate school of the Potsdam University and the HZB) and by the German Research Foundation (DFG) within the collaborative research center 951 "Hybrid Inorganic/Organic Systems for Opto-Electronics (HIOS)".

**Author contributions.**

**Competing financial interests.** The authors declare no competing financial interests.

**Data availability.** The data that support the plots within this paper and other findings of this study are available from the corresponding authors upon reasonable request.

## Supplementary Information

**The perovskite/transport layer interfaces dominate non-radiative recombination in efficient perovskite solar cells**

Martin Stolterfoht[1,*], Pietro Caprioglio[1,3], Christian M. Wolff[1], José A. Márquez[2], Joleik Nordmann[1], Shanshan Zhang[1], Daniel Rothhardt[1], Ulrich Hörmann[1], Alex Redinger[2], Lukas Kegelmann[3], Steve Albrecht[3], Thomas Kirchartz[4], Michael Saliba[5], Thomas Unold[2,*], Dieter Neher[1,*]

[1]Institute of Physics and Astronomy, University of Potsdam, Karl-Liebknecht-Str. 24-25, D-14476 Potsdam-Golm, Germany.

[2]Department of Structure and Dynamics of Energy Materials, Helmholtz-Zentrum-Berlin, Hahn-Meitner-Platz 1, D-14109 Berlin, Germany

[3]Young Investigator Group Perovskite Tandem Solar Cells, Helmholtz-Zentrum Berlin für Materialien und Energie GmbH, Kekuléstraße 5, 12489 Berlin, Germany

[4]Institut für Energie- und Klimaforschung, Forschungszentrum Jülich GmbH, 52425 Jülich, Germany

[5]Soft Matter Physics, Adolphe Merkle Institute, CH-1700 Fribourg, Switzerland

**E-mail:** stolterf@uni-potsdam.de, unold@helmholtz-berlin.de, neher@uni-potsdam.de

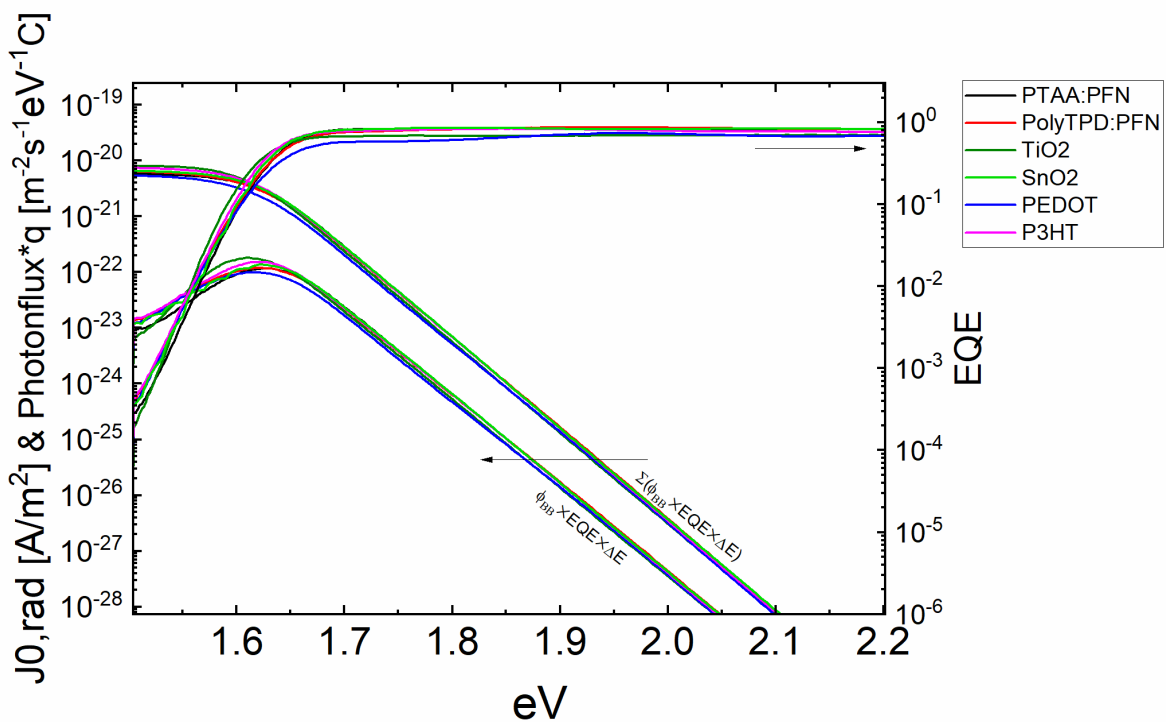

**Supplementary Figure S1.** External Quantum Efficiency (EQE) spectra, the product of the black body ($\phi_{BB}$) spectrum and the EQE, and the integral of $\phi_{BB} * \mathrm{EQE}$. The graphs shows that $J_{0,\mathrm{rad}}$ is very similar for all system ($6.5 \pm 1 \times 10^{-21}$ A/m$^2$) independent of the bottom charge transport layer. This also suggests that the opto-electronic quality of the perovskite layer is not significantly altered due to the different HTL underneath.

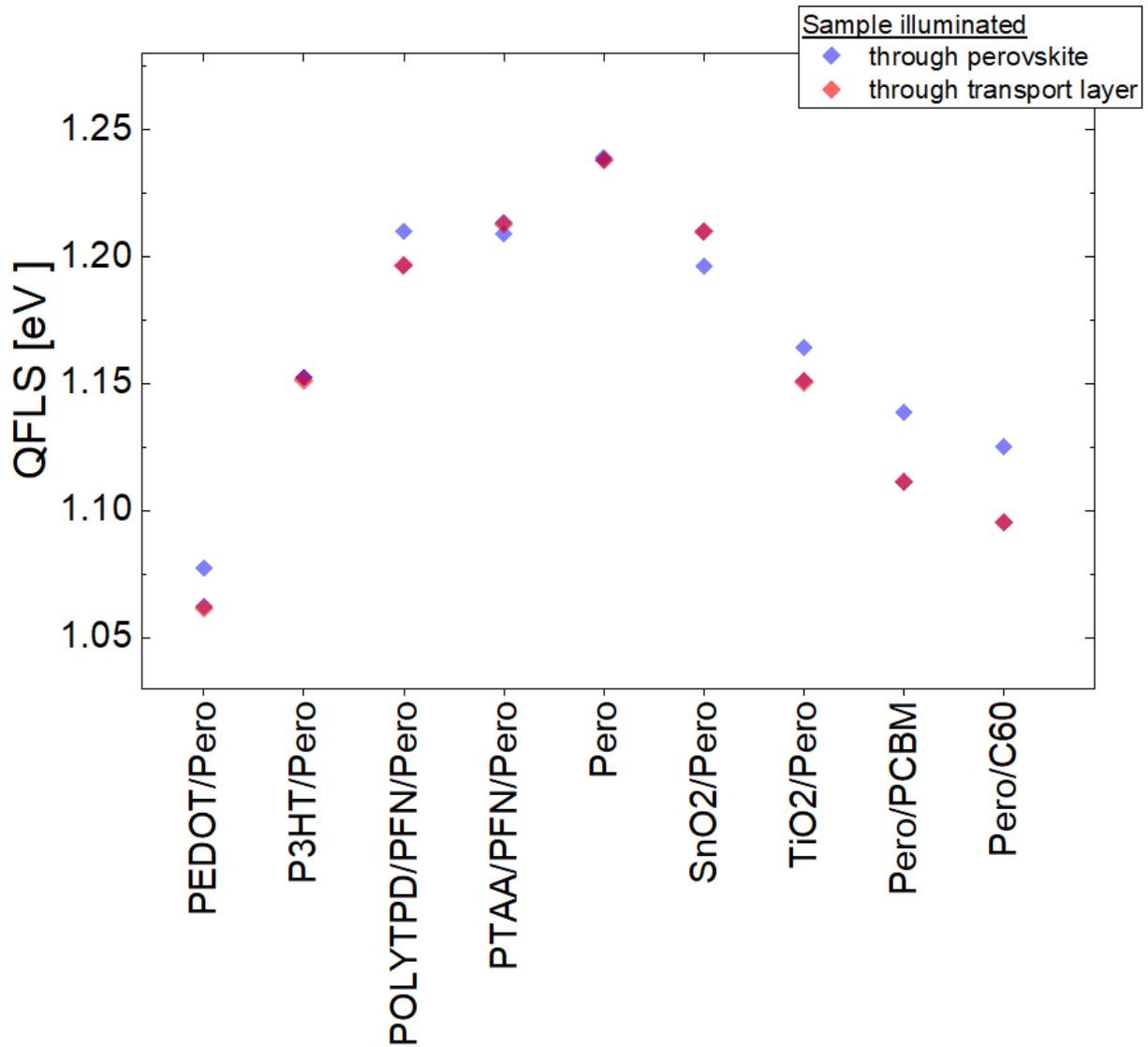

**Supplementary Figure S2.** Quasi-Fermi level splitting of various perovskite films illuminated through the perovskite (blue circles) or the charge transport layer (red symbols) using a 445 nm CW laser. In case of the neat perovskite film, the red symbol corresponds to a measurement through the bottom glass substrate. We note that the 445 nm laser is absorbed within a narrow window in the perovskite layer (<150 nm penetration depth) according to optical transfer matrix simulations which are also shown in **Supplementary Figure S12**. The graph shows that illuminating through the electron transport layers (ETLs) $C_{60}$, PCBM causes a significantly lower QFLS (up to 30 meV) compared to illumination through the perovskite, which is attributed to substantial parasitic absorption in the ETL at this wavelength. A smaller difference in the QFLS depending on the illumination side was observed for the other transport layers.

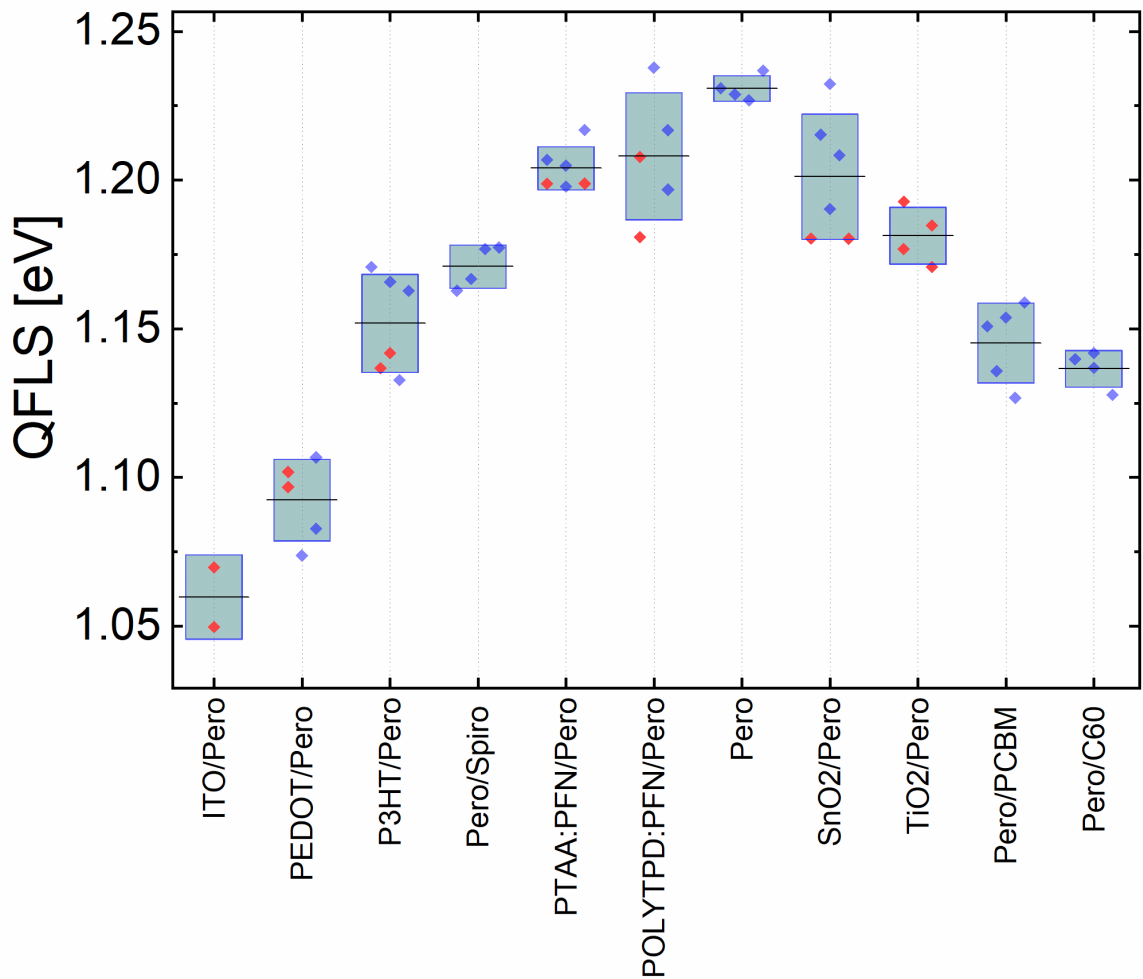

**Supplementary Figure S3.** The obtained quasi-Fermi level splitting of perovskite films including the studied hole and electron transporting materials and the neat absorber layer. Each data point corresponds to a different sample film. For each film an area of 1 cm$^2$ was illuminated and the average QFLS plotted. We also studied films on glass and glass/ITO substrates (glass/FTO in case of TiO2) which are more relevant for actual devices. The values obtained on glass/ITO (glass/FTO) are plotted in red (films on glass in blue). We note small differences between these two substrates indicating small losses between the HTL and the metal electrode in some cases. The lines show the mean values and the boxes the standard deviations.

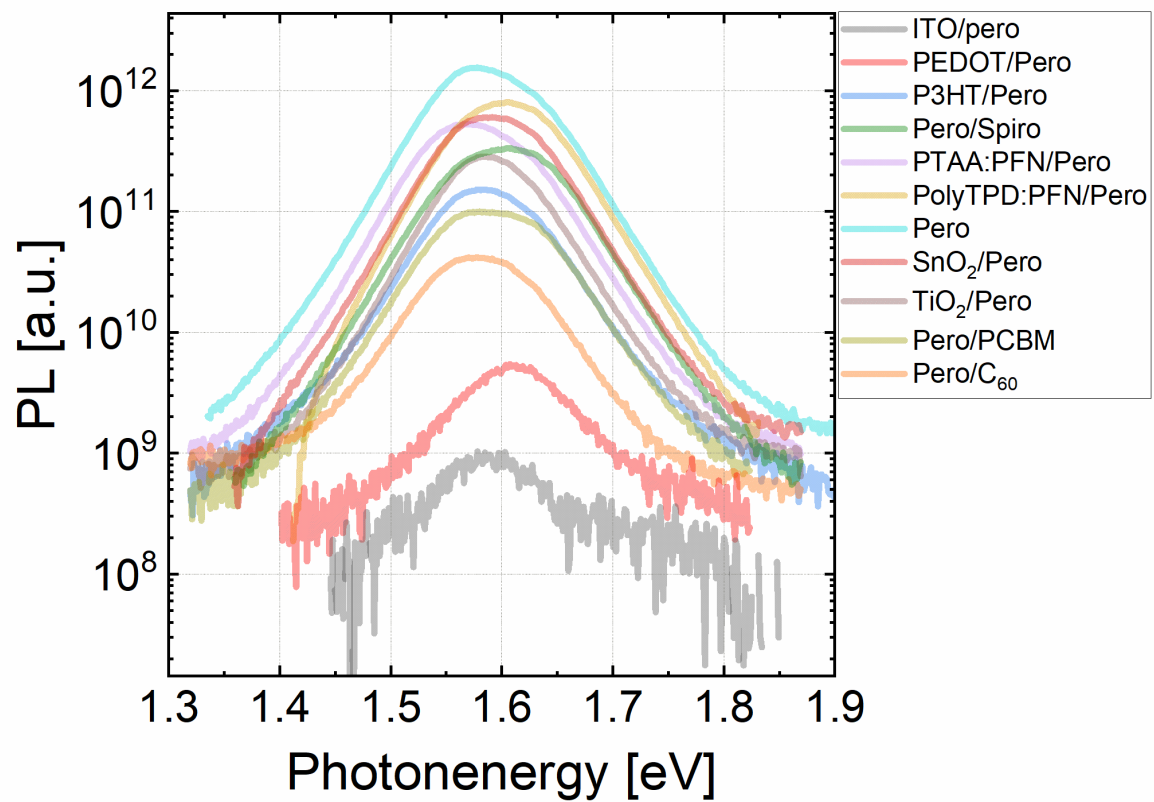

**Supplementary Figure S4.** Representative PL spectra of the bare perovskite film and perovskite films with different electron and hole transport layers.

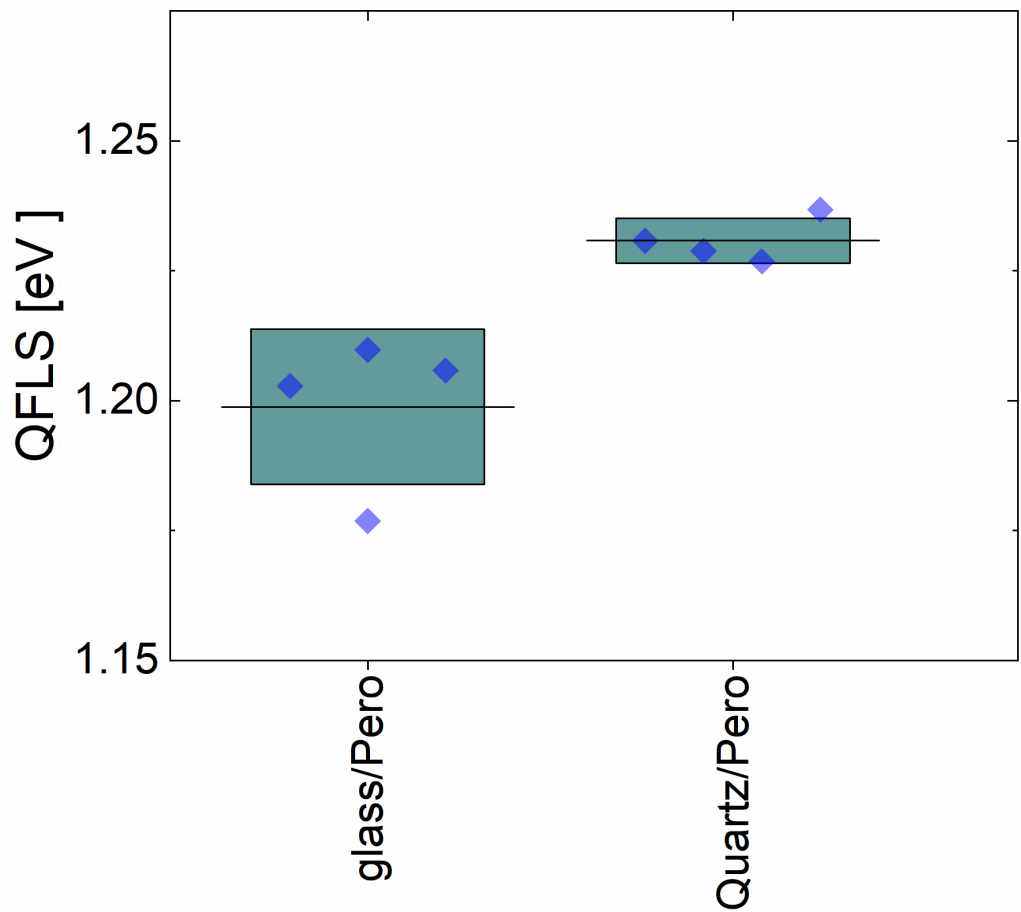

**Supplementary Figure S5.** The QFLS of triple cation perovskite films on glass and fused slilica shows that the latter substrate causes less non-radiative recombination losses, which indicates some recombination is occurring at the glass/perovskite interface. The lines show the mean values and the boxes the standard deviations.

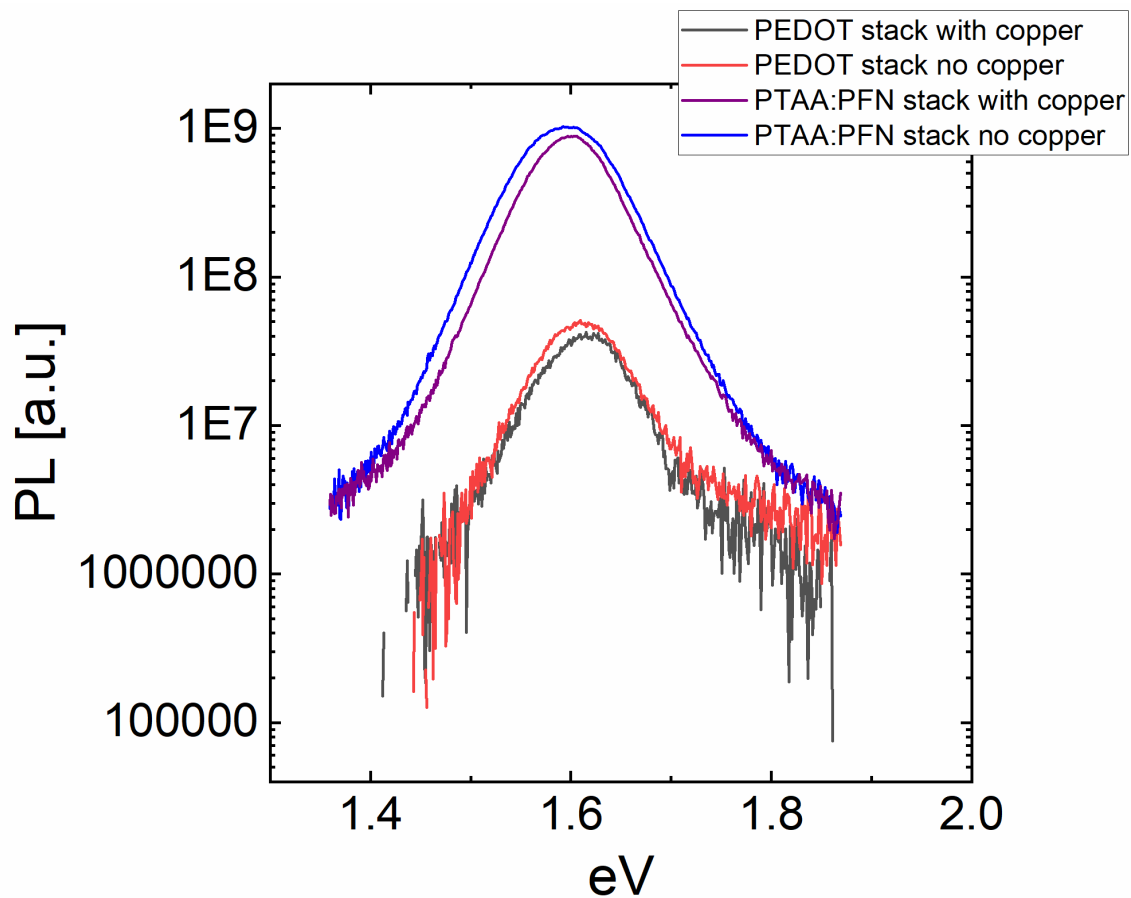

**Supplementary Figure S6.** Quasi-Fermi level splitting obtained on glass/ITO/PEDOT/pero/C60 and glass/ITO/PTAA/PFN/pero/C60 stacks with and without the copper electrode. No significant difference in the emission is observed in the presence of copper which is attributed to the fact that all samples are placed on a reflective sample holder in the Ulbricht sphere where the PL experiment was performed. Thus, emission that is emitted to the bottom sample holder is likely reflected back, similar to light that is emitted to the copper electrode which might explain the small impact of the copper electrode.

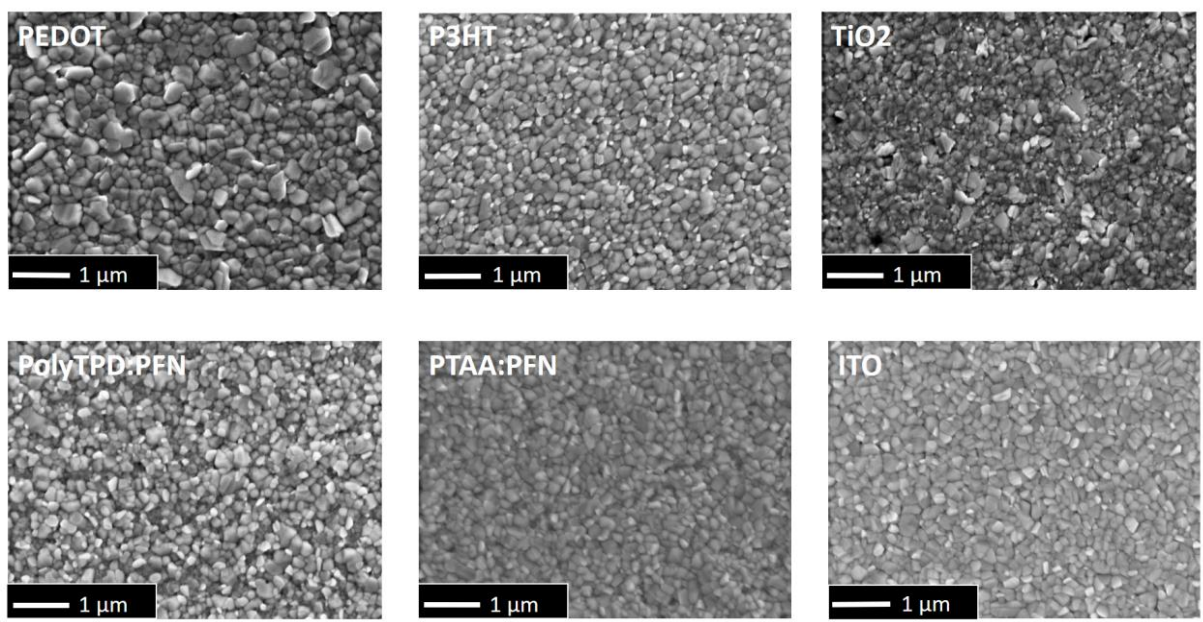

**Supplementary Figure S7.** Scanning electron microscopy (SEM) top sectional images of perovskite films fabricated on different underlying charge transport layers reveal differences in the perovskite morphology. Remarkably, are the substantially larger grains on PEDOT hole transport layers (despite their low radiative efficiency) and the broad distribution of different grain sizes on TiO$_2$ films. Relatively small grains are observed on PolyTPD:PFN, ITO, P3HT and PTAA:PFN bottoms layers. Overall, no clear correlation between the perovskite morphology and the photovoltaic performance can be deduced from these SEM results.

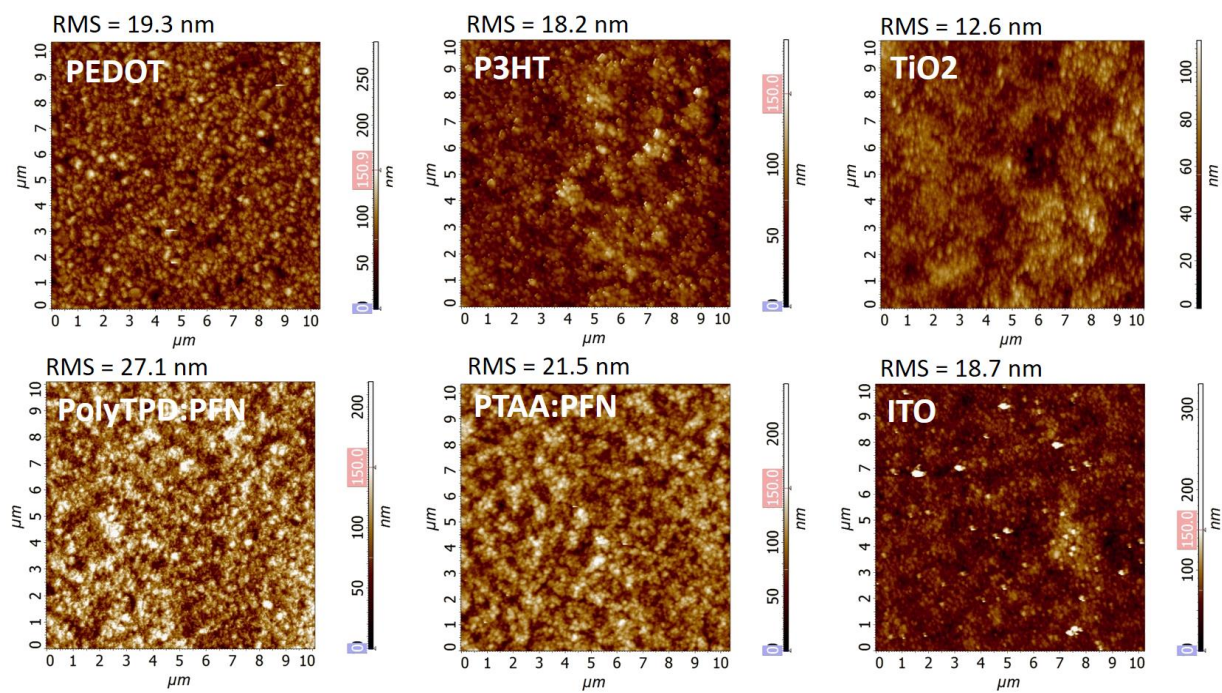

**Supplementary Figure S8.** Atomic Force Microscopy (AFM) top sectional measurements on perovskite films fabricated on different underlying charge transport layers reveal differences in the root mean square roughness (RMS) for each layer. Interestingly, the most efficient films in terms of photoluminescence exhibit a slightly rougher surface compared to the others, while films on TiO$_2$ were the smoothest.

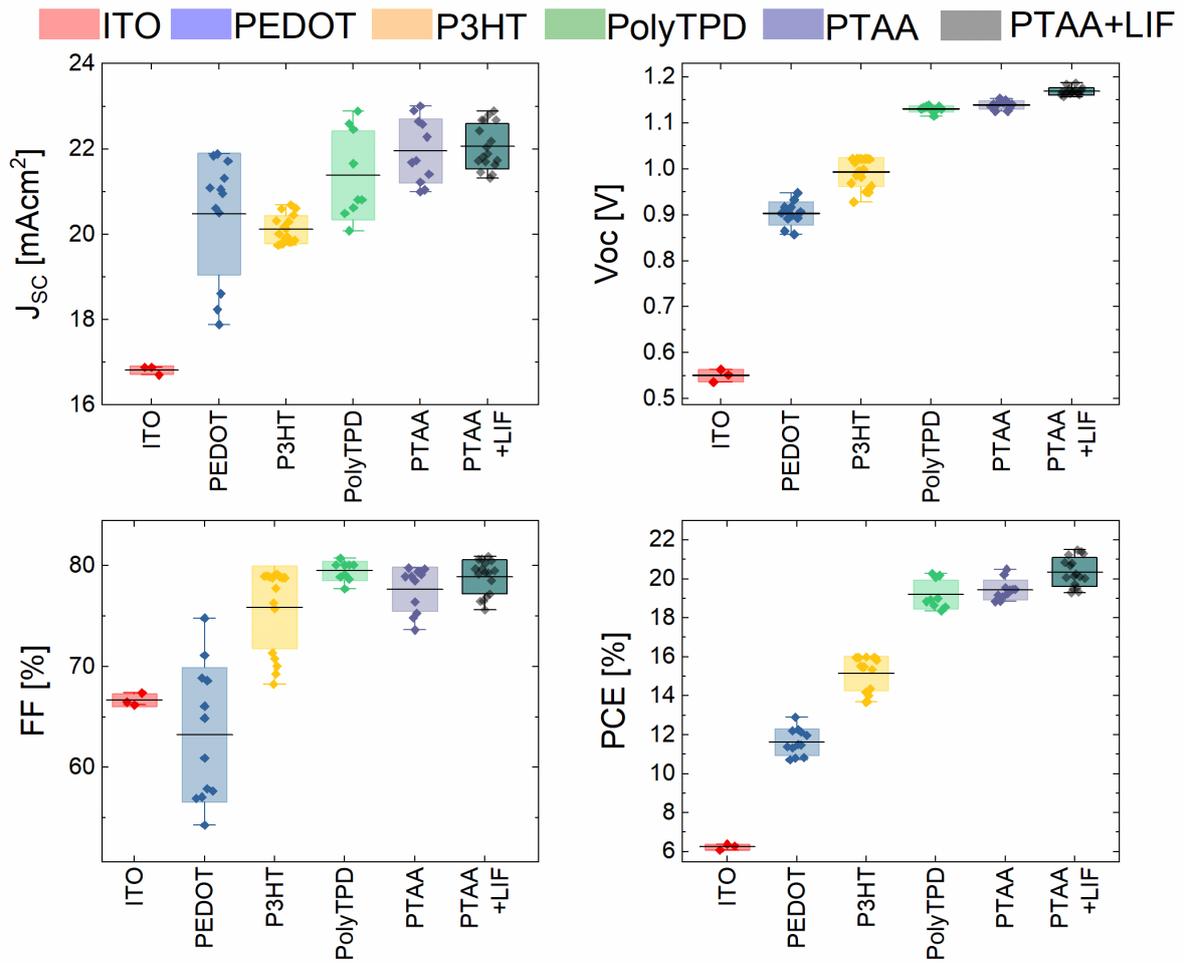

**Supplementary Figure S9.** Device statistics of 6 mm$^2$-size perovskite solar cells (with standard configuration ITO/HTL/perovskite/C60/BCP/Cu) showing the impact of the hole transport layer on the solar cell parameters. The average $V_{OC}$ values are plotted in Figure 3. The cells plotted on the right in each panel were fabricated using and additional LiF layer (~1 nm) between the perovskite and C$_{60}$ which allowed efficiencies above 20%. The lines show the mean values and the boxes the standard deviations.

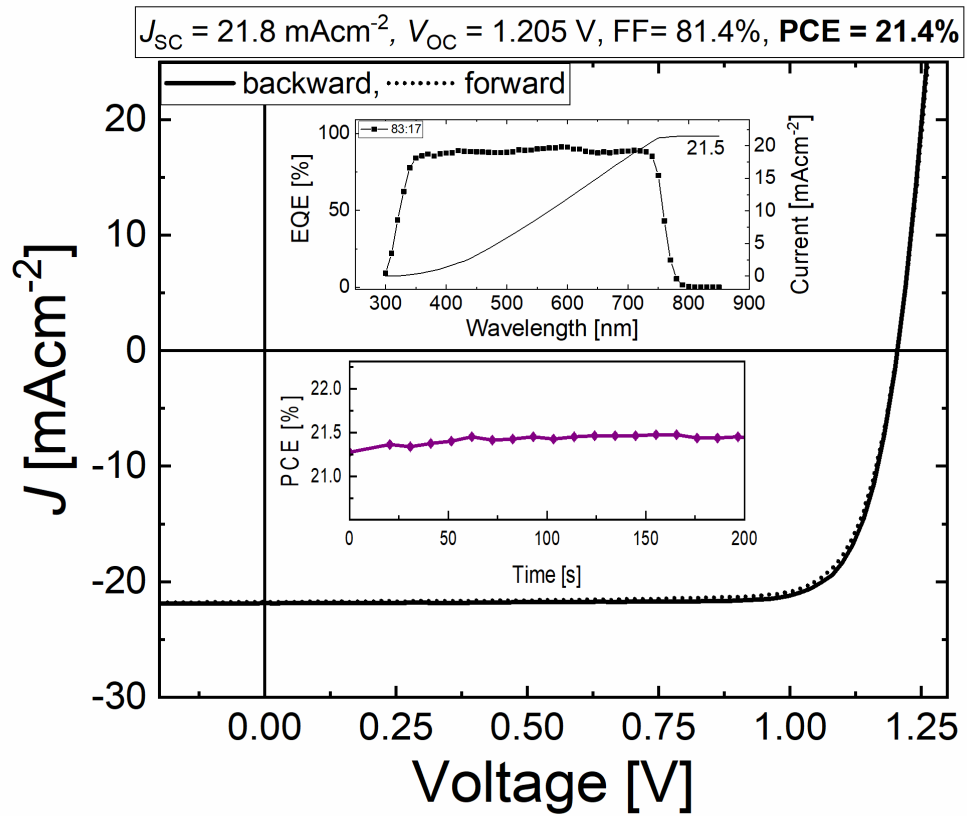

**Supplementary Figure S10.** *JV-characteristics of one of our most efficient cells fabricated at low temperatures (100 °C) using the standard $Cs_{0.05}(FA_{0.83}MA_{0.17})_{0.95}PbI_{0.83}Br_{0.17}$ triple cation perovskite absorber with a bandgap of approximately 1.6 eV, with PTAA/PFN and LiF/C60 as hole-and electron selective CTLs. The inset shows the stabilized efficiency evolution of the cell and the external quantum efficiency spectrum. The integrated product of the EQE and the solar spectrum (21.5 mAcm$^{-2}$) closely matches the measured short-circuit current density under the solar simulator (21.8 mAcm$^{-2}$).*

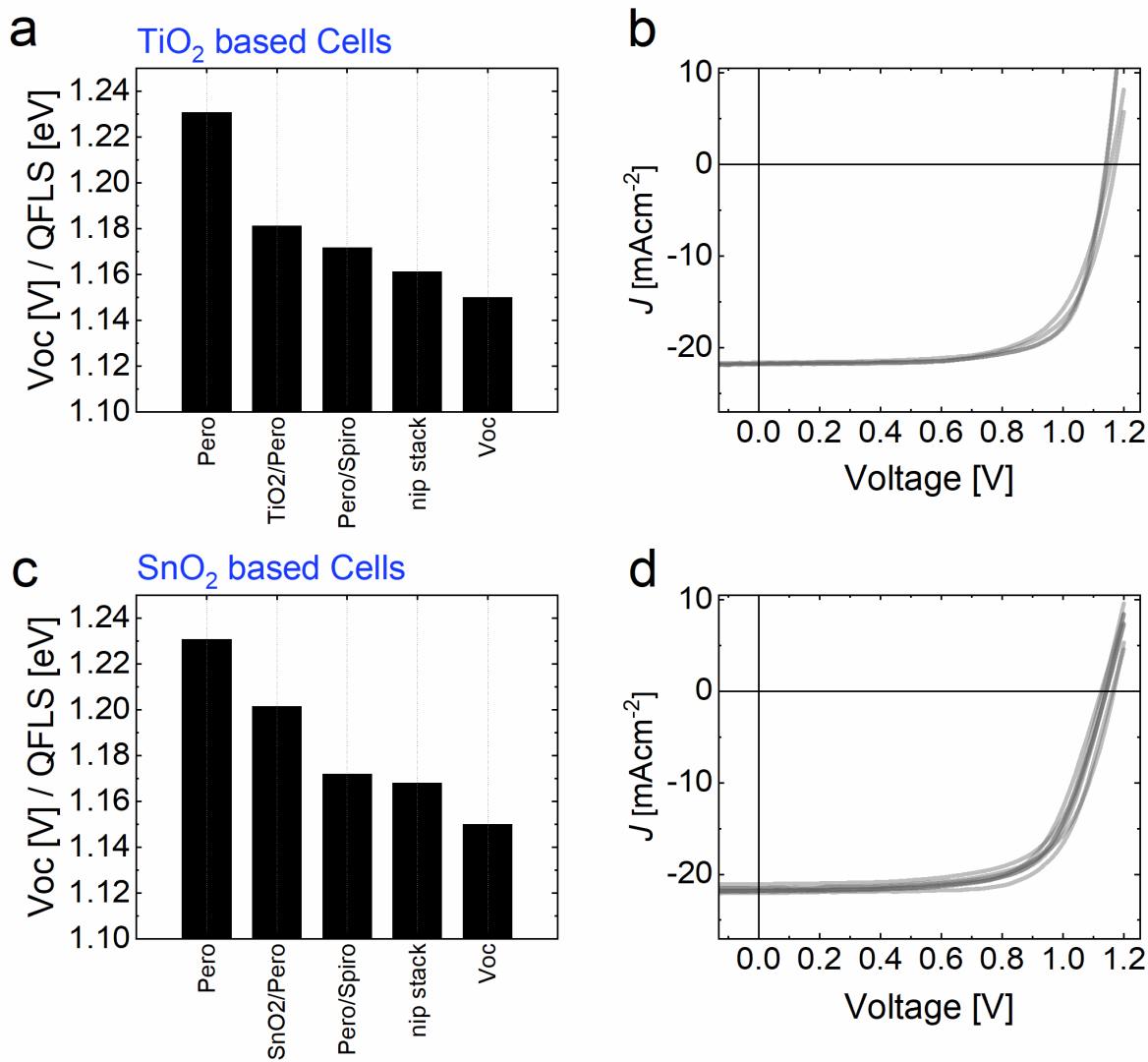

**Supplementary Figure S11.** Quasi-Fermi level splitting of the individual perovskite/transport layer films as well as the average device $V_{OC}$ of *nip* cells based on (**a, b**) TiO2 and (**c, d**) SnO₂ confirming that substantial interfacial non-radiative recombination losses lower the QFLS of the perovskite (1.23 eV) to 1.16-1.17 eV in the stack. For both cell types, the non-radiative recombination losses at the perovskite/spiroOMeTAD junction appear to limit the QFLS of the complete stack. For TiO₂ cells, 2 substrates with 4 pixels (30 mm²) in total were fabricated of which the *JV*-curves are shown in panel (**b**) with efficiencies of around 19%. For SnO₂ cells, 2 substrates with 12 pixels in total (16 mm²) were fabricated of which the *JV*-curves are shown in panel (**d**) with efficiencies up to 18% (max).

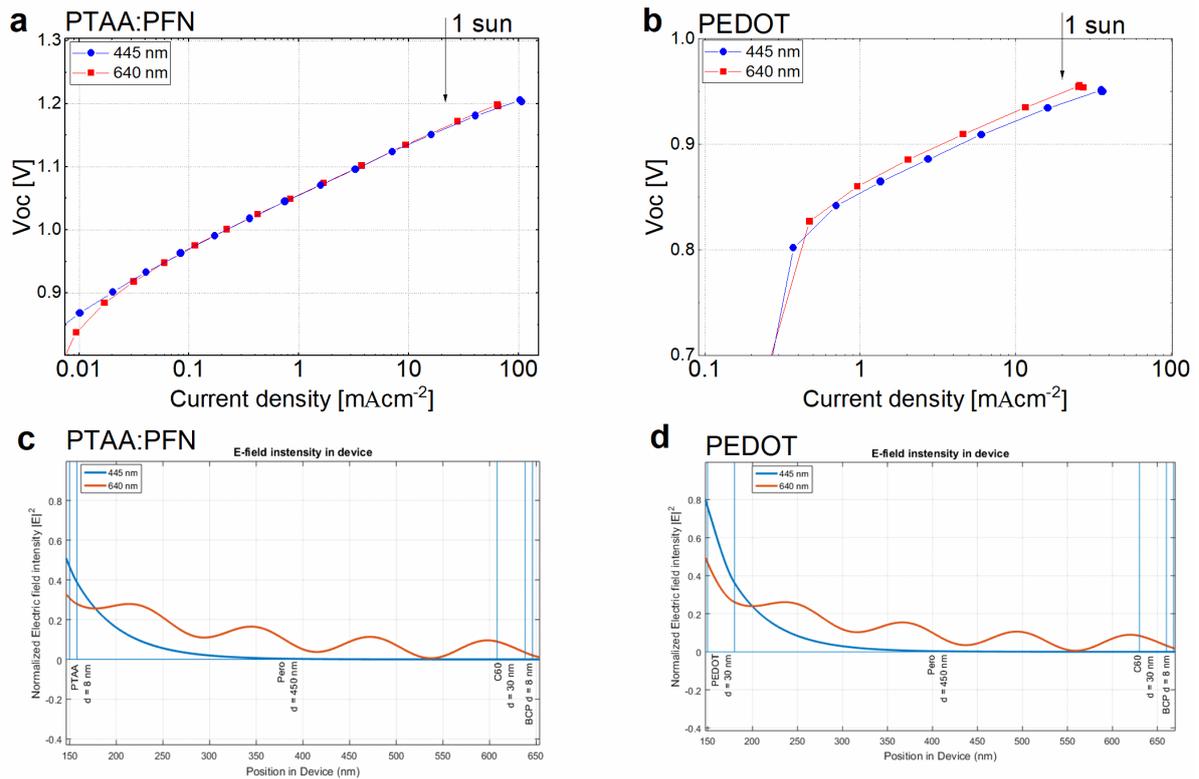

**Supplementary Figure S12.** $V_{OC}$ vs. short-circuit current for two different laser wavelengths (445 nm and 638 nm) on *pin*-type devices with (**a**) PTAA:PFN and (**b**) PEDOT as hole transport layer, while C$_{60}$ was used as electron transport layer in both cases. The graph demonstrates that the $V_{OC}$ is essentially independent on the initial carrier generation profile over several orders of magnitude in laser intensity (or short-circuit current ). (**c**) and (**d**) show the corresponding *E*-field intensity in the two devices which was simulated based on optical transfer matrix simulations using an open source code developed by McGehee et al. which was adapted from [*J. Appl. Phys.*, 86, 487 (1999); *J. Appl. Phys.*, 93, 3693 (2003)].

**Supplementary Table S1.** SCAPS simulation parameters for the simulations shown in **Figure 4**.

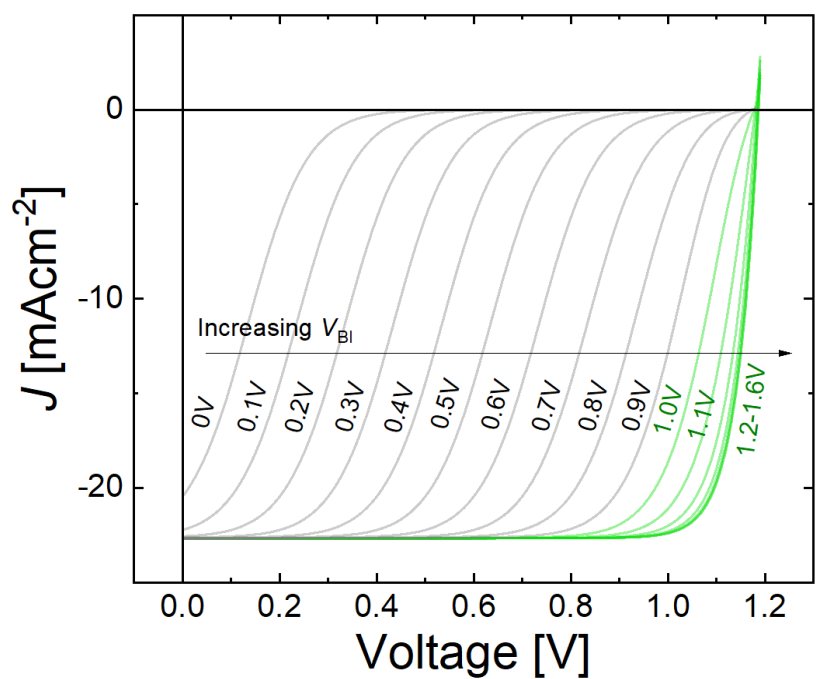

**Supplementary Figure S13.** Simulated *JV*-curves for different built-in voltages ($V_{BI}$) across the device. The $V_{BI}$ is given by the workfunction difference of the electrodes and is limited by the perovskite bandgap of 1.6 eV. The $V_{BI}$ was varied by equally reducing the energetic offsets between and the perovskite valence/conduction bands and the workfunctions of the metals at the bottom and top contact, respectively. The results suggest that a considerable $V_{BI}$ of ≈1V is required in order to efficiently extract the charges and reproduce experimental *JV*-curves.

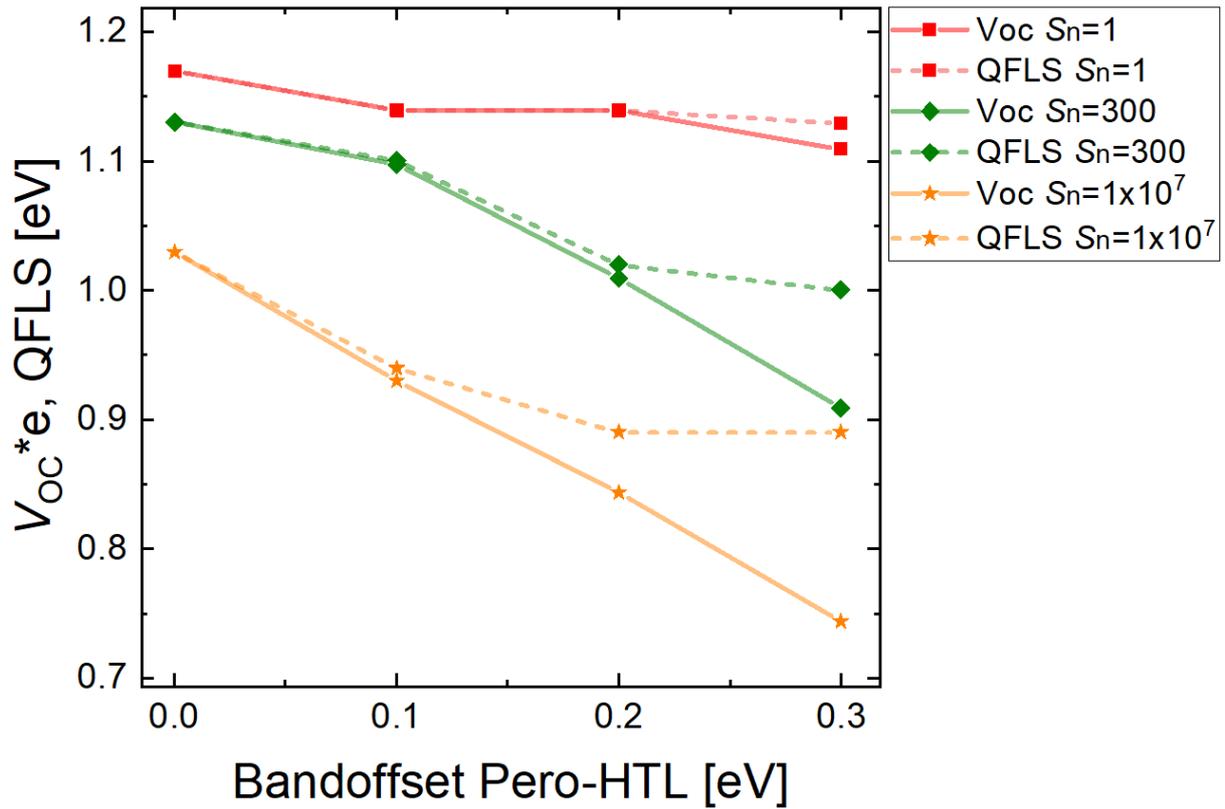

**Supplementary Figure S14.** Device simulations of $V_{OC}$ and average quasi-Fermi level splitting (QFLS) on PTAA/perovskite/C60 stacks for different interface recombination velocities of electrons at the HTL/perovskite interface ($S_n$) to reveal the origin of a mismatched QFLS and device $V_{OC}$. The simulations predict such a mismatch in case of a majority carrier band offset ($E_{maj}$) between the perovskite valence band and the highest occupied molecular orbital of PTAA if the interface recombination velocities are above 1 cm/s. Notably, based on these simulations we expect no QFLS-$V_{OC}$ mismatch in absence of a band offset regardless of the interface recombination velocity. The simulated $V_{OC}$ for the most realistic scenario with $S_n$ = 300 cm/s (green curve) at the HTL/perovskite interface shows that that even small majority carrier band offsets larger than >0.1 eV are already inconsistent with the experimentally measured $V_{OC}$'s of ~1.14 V in the PTAA/PFN/perovskite/C$_{60}$ device. We also note that for low interface recombination velocities $S_n$~1cm/s, the simulated QFLS and $V_{OC}$ are limited by the interface recombination velocity $S_p$ at the perovskite/ETL interface which was set to 1000 cm/s for these simulations. Voltages above 1.26 V can be achieved in the limit of negligible recombination at both interfaces (< 1 cm/s).

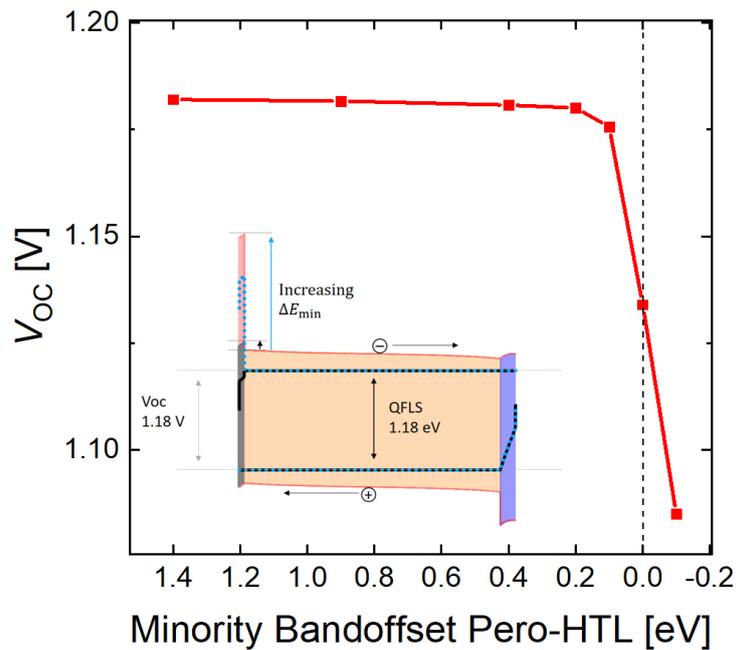

**Supplementary Figure S15.** Simulated open-circuit voltage versus minority carrier band offset ($E_{min}$) demonstrating that even smallest $\Delta E_{min}$ values of only 0.1 eV are in principal sufficient to prevent substantial charge recombination at the wrong interfaces. The primary reason for this result is that rapid recombination at the HTL/perovskite junction (which was set here to 200 cm/s) prevents minority carriers from entering the wrong CTL. The inset illustrates the energy bands for two device simulations where $\Delta E_{min}$ was increased from 0.1 eV to 1.4 eV demonstrating the nearly identical QFLS in the bulk (QFLS and energy bands are superimposed). The simulated electron and hole quasi-Fermi levels are shown by black lines and blue dots for $\Delta E_{min}$ energy offsets of 0.1 eV and 1.4 eV, respectively. Also shown are the conduction and valence bands in red, the device $V_{OC}$ and average quasi-Fermi level splitting (QFLS). We acknowledge that these simulations represent an ideal scenario where we only varied the position of the LUMO level of the HTL, however in reality such small minority carrier offsets would likely influence other critical parameters such as the accessible defect density for minority carriers which could, for example, cause much higher recombination velocities ($S$). Thus, while we can say that even small energetic offsets are sufficient to prevent minority carriers from entering the wrong contact, we cannot exclude that in reality such small offsets would cause much larger recombination losses by affecting other important parameters.